\documentclass[journal]{IEEEtran}

\ifCLASSINFOpdf
\else
\fi

\usepackage{hyperref}
\usepackage{amssymb}
\usepackage{amsmath}
\usepackage{graphicx}
\usepackage{color}
\usepackage{multirow}
\usepackage{tabto}
\usepackage{float}

\let\labelindent\relax
\usepackage{enumitem}

\usepackage[T1]{fontenc}
\usepackage{comment}

\usepackage[caption=false]{subfig}

\usepackage{threeparttable}

\usepackage[table,xcdraw]{xcolor}
\usepackage[ruled, vlined, linesnumbered]{algorithm2e}

\usepackage{etoolbox}
\appto\TPTnoteSettings{\scriptsize}
\usepackage{cite}

\usepackage[switch,pagewise]{lineno}

\begin{document}

\twocolumn

%
\title{Implementation and Analysis of QUIC for MQTT}

\author{{Puneet Kumar and Behnam Dezfouli}\\
\IEEEauthorblockA{Internet of Things Research Lab, Department of Computer Engineering, Santa Clara University, USA
\\
\texttt{\{pkumar,bdezfouli\}@scu.edu }
}}

\maketitle

\begin{abstract}
Transport and security protocols are essential to ensure reliable and secure communication between two parties. 
For IoT applications, these protocols must be lightweight, since IoT devices are usually resource constrained. 
Unfortunately, the existing transport and security protocols -- namely TCP/TLS and UDP/DTLS -- fall short in terms of connection overhead, latency, and connection migration when used in IoT applications. 
In this paper, after studying the root causes of these shortcomings, we show how utilizing QUIC in IoT scenarios results in a higher performance. 
Based on these observations, and given the popularity of MQTT as an IoT application layer protocol, we integrate MQTT with QUIC.
By presenting the main APIs and functions developed, we explain how connection establishment and message exchange functionalities work.
We evaluate the performance of MQTTw/QUIC versus MQTTw/TCP using wired, wireless, and long-distance testbeds. 
Our results show that MQTTw/QUIC reduces connection overhead in terms of the number of packets exchanged with the broker by up to 56\%. 
In addition, by eliminating half-open connections, MQTTw/QUIC reduces processor and memory usage by up to 83\% and 50\%, respectively.
Furthermore, by removing the head-of-line blocking problem, delivery latency is reduced by up to 55\%.
We also show that the throughput drops experienced by MQTTw/QUIC when a connection migration happens is considerably lower than that of MQTTw/TCP.

\end{abstract}

\IEEEpeerreviewmaketitle

\section{Introduction} \label{introLabel}
The Internet of Things is the enabler of many applications, such as the smart home, smart cities, remote medical monitoring, and industrial control, by connecting a large number of sensors and actuators to the Internet.
Existing studies predict that the number of connected devices will surpass 50 billion by 2020 \cite{CiscoPredict}.
To facilitate interconnection and software development, the communication between IoT devices usually employs a protocol stack similar to that of regular Internet-connected devices such as smartphones and laptops.
Specifically, IP (or 6LowPAN \cite{silva2009adaptation}) and transport layer protocols are provided by various protocol stacks (e.g., $\mu$IP \cite{dunkels2003full}, LwIP \cite{dunkels2002lwip}) to enable interconnectivity.


\subsection{TCP and UDP}
\label{intro_TCP_UDP}
The primary responsibility of the transport layer is to support exchanging segments between the two end-to-end communicating applications.
Among the transport layer protocols, TCP (Transport Layer Protocol) and UDP (User Datagram Protocol) are the most widely used, depending on the application at hand.
TCP provides a reliable end-to-end connection and implements congestion control mechanisms to avoid buffer overflow at the receiver.
During the past couple of decades, several improved versions of TCP have been proposed to address the increasing demand for throughput \cite{henderson2012newreno,xu2004binary}.
However, these features impose high overhead in terms of connection establishment and resource (i.e., processor, memory, energy) utilization.
UDP, on the other hand, does not provide any of the above-mentioned features and therefore, its overhead is significantly lower than that of TCP.

While throughput is the main performance metric for user traffic such as voice and video, the prevalent communication type of IoT, which is machine-to-machine (M2M), is characterized by short-lived bursts of exchanging small data chunks \cite{betzler2016coap,dezfouli2017rewimo}.
In addition, compared to user devices such as smartphones and laptops, IoT devices are usually resource constrained in terms of processing, memory, and energy \cite{shang2016challenges,dezfouli2018empiot}.
Using TCP in IoT domains to satisfy reliability and security requirements, therefore, imposes high overhead.
Specifically, the shortcomings of TCP when used in IoT applications are as follows:
\begin{itemize} [itemsep=0pt, topsep=0pt, leftmargin=*]
    \item [--] Connection startup latency is highly affected  by the TCP handshake. 
    This handshake requires 1 Round-Trip Time (RTT) for TCP and  2 or 3 RTTs when TLS (Transport Layer Security) is added to this protocol \cite{freierrfc}.
    The overhead impact is even higher in IoT scenarios where unreliable wireless links cause frequent connection drops \cite{atzori2010internet}.
    In these scenarios, imposing a high connection establishment overhead for the exchange of a small amount of data results in wasting the resources of devices.
    TCP Fast Open \cite{radhakrishnan2011tcp} seeks to address this problem by piggybacking data in SYN segments in repeated connections to the same server. 
    This solution is not scalable since the TCP SYN segment can only fit a limited amount of data \cite{langley2017quic}.
    %
    %
    \item [--] IoT devices are often mobile, and as such, supporting connection migration is an essential requirement \cite{soldatos2012convergence, distefano2012sensing,hussain2017seamblue,dezfouli2018review}.
    However, any change in network parameters (such as IP address or port) breaks the connection.
    In this case, either the connection must be re-established, or a gateway is required to reroute the data flow.
    Unfortunately, these solutions increase communication delay and overhead, which might not be acceptable in mission-critical applications such as medical monitoring \cite{dezfouli2017rewimo}. 
   \item [--] To preserve energy resources, IoT devices usually transition between sleep and awake states \cite{dezfouli2015modeling,nikoukar2018low}.
    In this case, a TCP connection cannot be kept open without employing \textit{keep-alive} packets.
    These keep-alive mechanisms, however, increase resource utilization and bandwidth consumption. 
    Without an external keep-alive mechanism, IoT devices are obliged to re-establish connections every time they wake from the sleep mode.     
    \item [--] In disastrous events such as unexpected reboots or a device crash, TCP connections between client and server might end up out of state.
    This undefined state is referred to as TCP \textit{half-open connections} \cite{saini2017evaluating}. 
    A half-open connection consumes resources such as memory and processor time. 
    In addition, it can impose serious threats such as SYN flooding \cite{chimkode2017design, ahmed2017defense}. 
    \item [--] If packets are dropped infrequently during a data flow, the receiver has to wait for dropped packets to be re-transmitted in order to complete the packet re-ordering.  
    This phenomena, which impedes packet delivery performance, is called the \textit{head-of-line blocking} \cite{nageswaraanalysis, bziuk2018http, scharf2006head}.     %
\end{itemize}

Despite the aforementioned shortcomings, several IoT application layer protocols rely on TCP, and some of them offer mechanisms to remedy these shortcomings.
For example, MQTT \cite{banks2014message} employs application layer keep-alive messages to keep the connection alive. 
This mechanism also enables MQTT to detect connection breakdown and release the resources. 

Another transport layer protocol used in IoT networks is UDP. 
Generally, UDP is suitable for applications where connection reliability is not essential. 
Although this is not acceptable in many IoT scenarios, several IoT application layer protocols rely on UDP due to its lower overhead compared to TCP.
These protocols usually include mechanisms to support reliability and block transmission (e.g., CoAP \cite{shelby2014ietf}).

\subsection{TLS and DTLS}
In addition to reliability, it is essential for IoT applications to employ cryptographic protocols to secure end-to-end data exchange over transport layer.
TLS \cite{dierks2008rfc} is the most common connection-oriented and stateful client-server cryptographic protocol. 
Symmetric encryption in TLS enables authenticated, confidential, and integrity-preserved communication between two devices.
The key for this symmetric encryption is generated during the TLS handshake and is unique per connection.
In order to establish a connection, the TLS handshake can require up to two round-trips between the server and client. 
However, since connections might be dropped due to phenomenons such as sleep phases, connection migration, and packet loss, the overhead of establishing secure connections imposes high overhead. 
In order to address this concern, a lighter version of TLS for datagrams, named DTLS (Datagram Transport Layer Security) \cite{rescorla2016datagram}, has been introduced. 
Unlike TLS, DTLS does not require a reliable transport protocol as it can encrypt or decrypt out-of-order packets. 
Therefore, it can be used with UDP. 
Although the security level offered by DTLS is equivalent to TLS, some of the technical differences include:
the adoption of stream ciphers is prohibited, and an explicit sequence number is included in every DTLS message. 
Compared to TLS, DTLS is more suitable for resource-constrained devices communicating through an unreliable channel.
However, similar to TLS, the underlying mechanism in DTLS has been primarily designed for point-to-point communication. 
This creates a challenge to secure one-to-many connections such as broadcasting and multicasting.
In addition, since DTLS identifies connections based on source IP and port number, it does not support connection migration \cite{DTLS_Prob_Connection_Migration}. 
Furthermore, DTLS handshake packets are large and may fragment each datagram into several DTLS records where each record is fit into an IP datagram.
This can potentially cause record overhead \cite{DTLS_record_overhead}.

\subsection{Contributions}
Given the shortcomings of UDP and TCP, we argue that the enhancement of transport layer protocols is a necessary step in the performance improvement of IoT applications. 
In order to address this concern, this paper presents the implementation and studies the integration of QUIC \cite{QUIC_Proto} with application layer to address these concerns. 
QUIC is a user space, UDP-based, stream-based, and multiplexed transport protocol developed by Google.
According to \cite{langley2017quic}, around 7\% of the world-wide Internet traffic employs QUIC. 
This protocol offers all the functionalities required to be considered a connection-oriented transport protocol. 
In addition, QUIC solves the numerous problems faced by other connection-oriented protocols such as TCP and SCTP \cite{stewart2007stream}. 
Specifically, the addressed problems are: reducing the connection setup overhead, supporting multiplexing, removing the head-of-line blocking, supporting connection migration, and eliminating TCP half-open connections. 
QUIC executes a cryptographic handshake that reduces the overhead of connection establishment by employing known server credentials learned from past connections. 
In addition, QUIC reduces transport layer overhead by multiplexing several connections into a single connection pipeline.
Furthermore, since QUIC uses UDP, it does not maintain connection status information in the transport layer.
This protocol also eradicates the head-of-line blocking delays by applying a lightweight data-structure abstraction called \textit{streams}.

At present, there is no open source or licensed version of MQTT using QUIC. 
Current MQTT implementations (such as Paho \cite{Paho_Eclipse}) rely on TCP/TLS to offer reliable and secure delivery of packets.
Given the potentials of QUIC and its suitability in IoT scenarios, in this paper we implement and study the integration of MQTT with QUIC.
First, since the data structures and message mechanisms of MQTT are intertwined with the built-in TCP and TLS APIs, it was necessary to redesign these data structures.
The second challenge was to establish IPC (Inter-Process Communication) between QUIC and MQTT.
MQTTw/TCP utilizes the available APIs for user space and kernel communication. 
However, there is no available API for QUIC and MQTT to communicate, as they are both user space processes.
To address these challenges, we have developed new APIs, which are referred to as \textit{agents}.
Specifically, we implemented two types of agents: server-agent and client-agent, where the former handles the broker-specific operations and the latter handles the functionalities of publisher and subscriber. 
This paper presents all the functions developed and explains the connection establishment and message exchange functionalities by presenting their respective algorithms.
The third challenge was to strip QUIC of mechanisms not necessary for IoT scenarios.
QUIC is a web traffic protocol composed of a heavy code footprint (1.5GB \cite{chromechromium}). 
We have significantly reduced the code footprint (to around 22MB) by eliminating non-IoT related code segments such as loop network and proxy backend support.

Three types of testbeds were used to evaluate the performance of MQTTw/QUIC versus MQTTw/TCP: wired, wireless, and long-distance.
Our results show that, in terms of the number of packets exchanged during the connection establishment phase, MQTTw/QUIC offers a 56.2\% improvement over MQTTw/TCP. 
By eliminating half-open connections, MQTTw/QUIC reduces processor and memory usage by up to 83.2\% and 50.3\%, respectively, compared to MQTTw/TCP.
Furthermore, by addressing the head-of-line blocking problem, MQTTw/QUIC reduces message delivery latency by 55.6\%, compared to MQTTw/TCP.
In terms of connection migration, the throughput drop experienced by MQTTw/QUIC is significantly lower than that of MQTTw/TCP.

The rest of this paper is organized as follows.
Section \ref{QUIC} explains the QUIC protocol along with its potential benefits in IoT applications.
The implementation of QUIC for MQTT is explained in Section \ref{implementation}.
Performance evaluation and experimentation results are given in Section \ref{result_final}.
Section \ref{relatedwork} overviews the existing studies on QUIC and IoT application layer protocols.
The paper is concluded in Section \ref{conclusion}.

\section{QUIC} \label{QUIC}

QUIC employs some of the basic mechanisms of TCP and TLS, while keeping UDP as its underlying transport layer protocol.
QUIC is in fact a combination of transport and security protocols by performing tasks including encryption, packet re-ordering, and retransmission. 
This section overviews the main functionalities of this protocol and justifies the importance of its adoption in the context of IoT.

\subsection{Connection Establishment} 
\label{quicSecurity} 
QUIC combines transport and secure layer handshakes to minimize the overhead and latency of connection establishment.
To this end, a dedicated reliable stream is provided for the cryptographic handshake. 
Figure \ref{fig:ConSetupQUIC}(a) and (b) show the packets exchanged during the 1-RTT and 0-RTT connection establishment phases, respectively.
Connection establishment works as follows:
\begin{figure}[t]
    \centering
    \includegraphics[width=0.95 \linewidth]{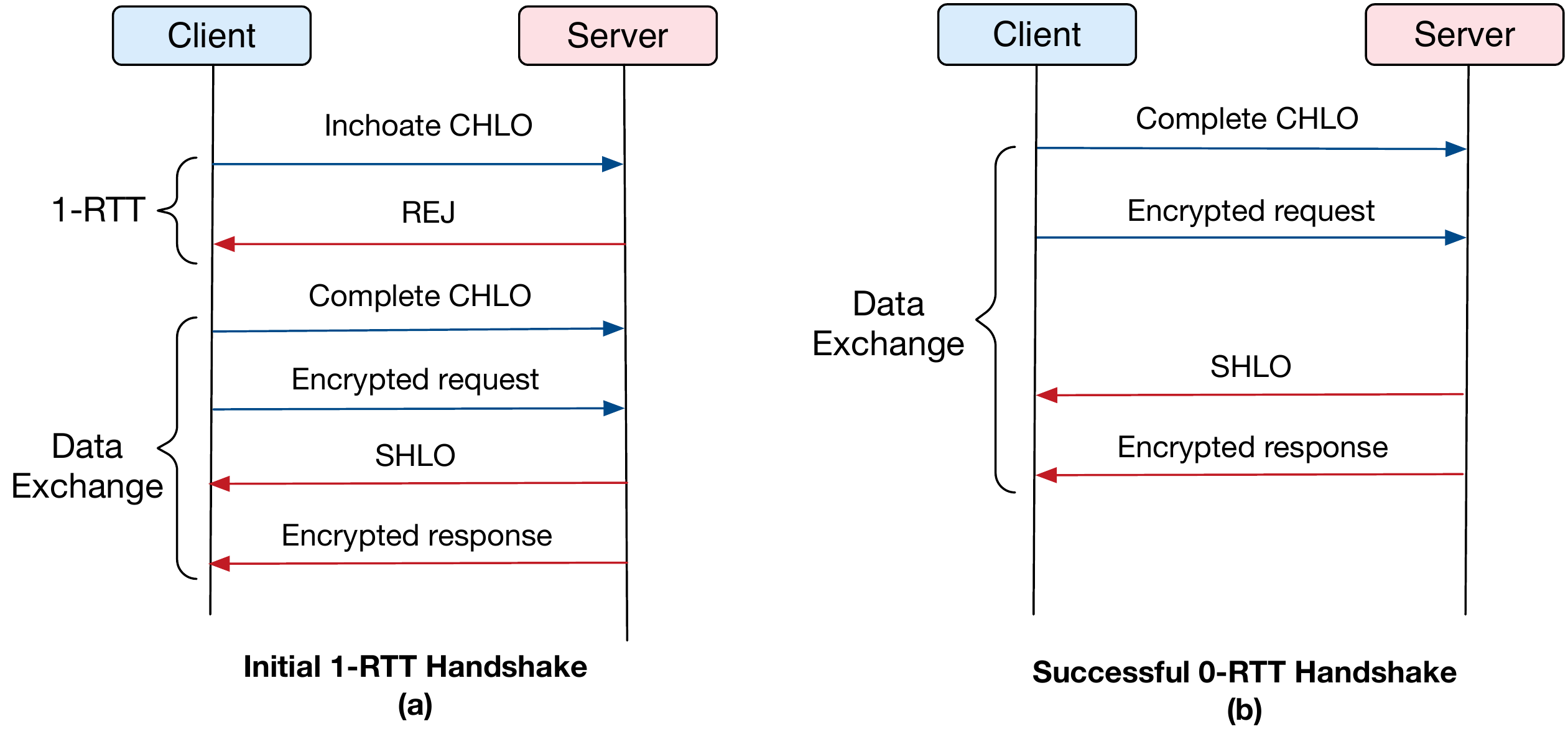}
    \captionsetup{font=scriptsize}
    \caption{The messages exchanged by QUIC during (a) 1-RTT \label{1RTTFig} and (b) 0-RTT connections. 
    \label{0RTTFig}
    }
    \label{fig:ConSetupQUIC}
\end{figure}
\begin{itemize} [leftmargin=*]
    \item [--] \textit{First handshake}. 
    In order to retrieve the server's configuration, the client sends an inchoate client hello (\texttt{CHLO}) message. 
    Since the server is an alien to the client, the server must send a \texttt{REJ} packet.
    This packet carries the server configuration including the long-term Diffie-Hellman value, 
    connection ID (\texttt{cid}),
    port numbers,  
    key agreement,
    and initial data exchange.
    After receiving the server's configuration, the client authenticates the server by verifying the certificate chain and the signature received in the \texttt{REJ} message.
    At this point, the client sends a complete \texttt{CHLO} packet to the server. 
    This message contains the client's ephemeral Diffie-Helman public value. 
    This concludes the first handshake.
    \item [--] \textit{Final and repeat handshake}. 
    After receiving the complete \texttt{CHLO} packet, the client has the initial keys for the connection and starts sending application data to the server. 
    For 0-RTT, the client must initiate sending encrypted data with its initial keys before waiting for a reply from the server. 
    If the handshake was successful, the server sends a server hello (\texttt{SHLO}) message.
    This concludes the final and repeat handshake. 

\end{itemize}

\begin{figure}[t]
    \centering
    \includegraphics[width=0.9\linewidth]{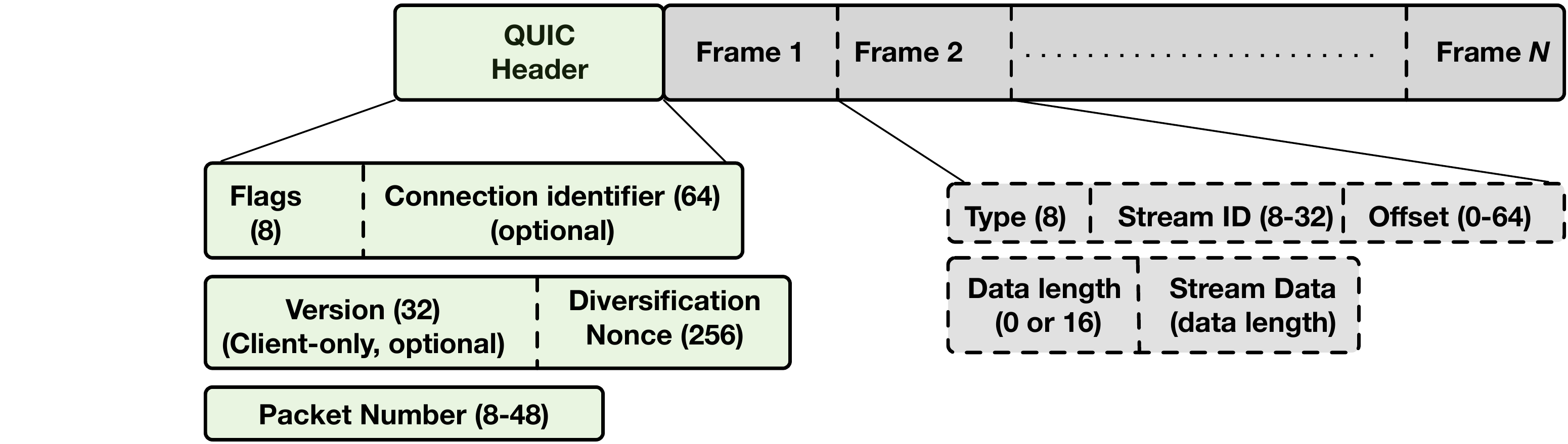}
    \captionsetup{font=scriptsize}
    \caption{{The solid and dashed lines show the clear-text and encrypted parts of a QUIC packet, respectively. 
    The non-encrypted part is used for routing and decrypting the encrypted part of the packet.} }
    \label{fig:QUICPacket}
\end{figure}

Except some handshake and reset packets, QUIC packets are fully authenticated and partially encrypted. 
Figure \ref{fig:QUICPacket} shows the non-encrypted and encrypted parts of the packet using solid and dotted lines, respectively.
The non-encrypted packet header is used for routing and decrypting the packet content. 
The flags encode the presence of \texttt{cid} and the length of the Packet Number (PN) field, which are visible to read the subsequent fields.

%
%

\subsection{Connection Migration} \label{CID}
The QUIC connections are identified by a randomly generated 64-bit Connection Identifier ({\texttt{cid}}).
A \texttt{cid} is allocated per connection and allows the clients to roam between networks without being affected by the changes in the network or transport layer parameters.
As shown in Figure \ref{fig:QUICPacket}, \texttt{cid} resides in the header (non-encrypted part) and makes the clients independent of network address translation (NAT) and restoration of connections.
The \texttt{cid} plays an important role in routing, specifically for connection identification purposes. 
Furthermore, using \texttt{cid}s enables multipath by probing a new path for connection. 
This process is called \textit{path validation} \cite{iyengar2017quic}. 
During a connection migration, the end point assumes that the peer is willing to accept packets at its current address. 
Therefore, an end point can migrate to a new IP address without first validating the peer's IP address. 
It is possible that the new path does not support the current sending rate of the endpoint. 
In this case, the end point needs to reconstitute its congestion controller \cite{CongestionControllerReset}. 
On the other hand, receiving non-probe packets \cite{NonProbePackets} from a new peer address confirms that the peer has migrated to the new IP address.

\subsection{Security}
For transport layer encryption, MQTTw/TCP usually relies on TLS/SSL.
The primary reasons why TLS/SSL cannot be used in QUIC were described in \cite{krawczyk2013security}.
In short, the TLS security model uses one session key, while QUIC uses two session keys. 
This difference, in particular, enables QUIC to offer 0-RTT because data can be encrypted before the final key is set. 
Thus, the model has to deal with data exchange under multiple session keys \cite{fischlin2014multi}. 
Therefore, MQTTw/QUIC uses its own encryption algorithm named QUIC Crypto \cite{Quic_Cry}.
This algorithm decrypts packets independently to avoid serialized decoding dependencies.
The signature algorithms supported by Crypto are ECDSA-SHA256 and RSA-PSS-SHA256.


\subsection{Multiplexing} 
\label{multiplexing}
Unlike TCP, QUIC is adept in transport layer header compression by using multiplexing. 
Instead of opening multiple connections from the same client, QUIC opens several streams multiplexed over a single connection.
Each connection is identified by a unique \texttt{cid}. 
The odd \texttt{cid}s are for client-initiated streams and even \texttt{cid}s are for server-initiated streams.
A stream is a lightweight abstraction that provides a reliable bidirectional byte-stream.
A QUIC stream can form an application message up to $2^{64}$ bytes.
Furthermore, in the case of packet loss, the application is not prevented from processing subsequent packets.
Multiplexing is useful in IoT applications where a large amount of data transfer is required per transaction. 
For example, this feature enhances performance for remote updates and industrial automation \cite{gruner2016restful}.

\subsection{Flow and Congestion Control} 
\label{congestion_control}

Similar to TCP, QUIC implements a flow control mechanism to prevent the receiver's buffer from being inundated with data \cite{langley2017quic}.
A slow TCP draining stream can consume the entire receiver buffer. 
This can eventually block the sender from sending any data through the other streams.
QUIC eliminates this problem by applying two levels of flow control: 
(i) Connection level flow control: limits the aggregate buffer that a sender can consume across all the streams on a receiver. 
(ii) Stream level flow control: limits the buffer per stream level. 
A QUIC receiver communicates the capability of receiving data by periodically advertising the absolute byte offset per stream in \textit{window update frames} for sent, received, and delivered packets.

QUIC incorporates a pluggable congestion control algorithm and provides a richer set of information than TCP \cite{QUIC_General}.
For example, each packet (original or re-transmitted) carries a new Packet Number (\texttt{PN}). 
This enables the sender to distinguish between the re-transmitted and original ACKs, hence removing TCP's re-transmission \textit{ambiguity problem}. 
QUIC utilizes a NACK based mechanism, where two types of packets are reported:
the largest observed packet number, and the unseen packets with a packet number lesser than that of the largest observed packet. 
A receive timestamp is also included in every newly-acked ACK frame.
QUIC's ACKs can also provide the delay between the receipt of a packet and its acknowledgement, which helps in calculating RTT. 
QUIC's ACK frames support up to 256 NACK ranges in opposed to the TCP's 3 NACK range \cite{iyengar2017quicCongestion}. 
This makes QUIC more resilient to packet reordering than TCP (with SACK).
The congestion control algorithm of QUIC is based on TCP Reno to determine the pacing rate and congestion window size \cite{iyengar2017quicCongestion}.
In addition, QUIC supports two congestion control algorithms: 
(i) Pacing Based Congestion Control Algorithm (PBCCA) \cite{aggarwal2000understanding}, and 
(ii) TCP CUBIC \cite{rhee2018cubic}.
The superior performance of QUIC's flow control over TCP for HTTP traffic has been demonstrated in the literature \cite{kakhki2017taking}.

\section{Integration and Implementation\\ of MQTT with QUIC}
\label{implementation}
This section presents the integration of MQTT with QUIC and is divided into six sub-sections. 
The first sub-section overviews the system architecture.
The second sub-section describes the definitions, methods, and assumptions.
The third and fourth sub-sections explain the operations of the APIs and functions developed for the broker and clients, respectively.
We present the common APIs and functions, which are used by the broker and client in the fifth sub-section.
A short discussion about code reduction is presented in the sixth sub-section.
In order to simplify the discussions, we refer to the publisher and the subscriber as \textit{client}.
The presented implementation for the client and broker are based on the open-source Eclipse Paho and Mosquitto \cite{Paho_Eclipse}, respectively.

\subsection{Architecture}
\begin{figure}[t]
    \centering
    \includegraphics[width=1.0\linewidth]{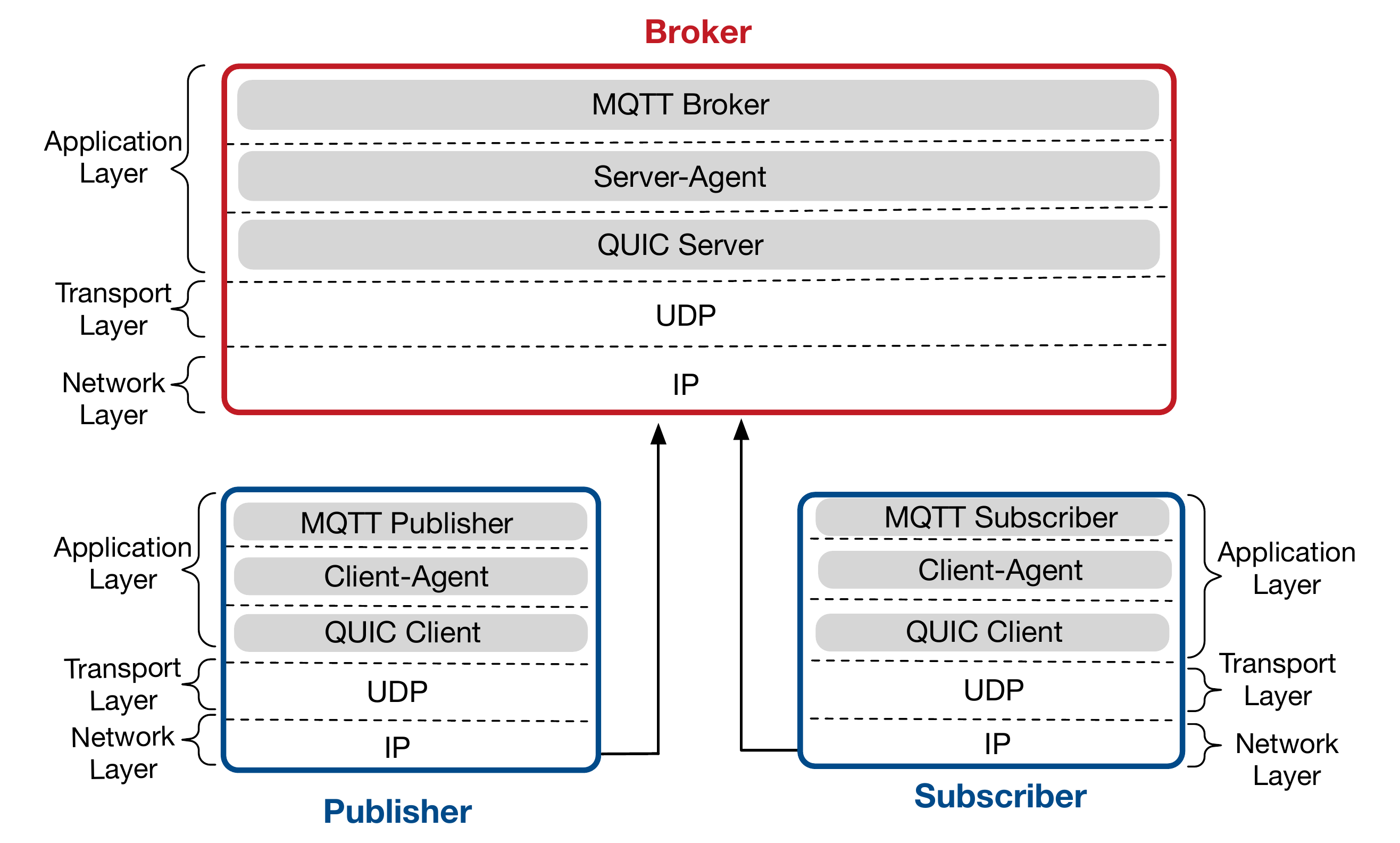}
    \captionsetup{font=scriptsize}
    \caption{{The high-level architecture of the proposed implementation.}}
    \label{fig:NetworkingStackAgents}
\end{figure}

Both MQTT and QUIC belong to the application layer.
To streamline their integration, either the QUIC library (\texttt{ngtcp2} \cite{ngtcp2}) must be imported into MQTT, or new interfaces must be created.
However, the former approach is not suitable for resource-constrained IoT devices as QUIC libraries are developed mainly for HTTP/HTTPS traffic, and therefore, impose a heavy code footprint.
In our implementation, we chose the latter approach and built a customized broker and client interfaces for MQTT and QUIC.
We refer to these interfaces as \textit{agents}, which enable IPC between the MQTT and QUIC.
Figure \ref{fig:NetworkingStackAgents} shows a high level view of the protocol stack architecture.

Since QUIC uses UDP, connection-oriented features such as reliability, congestion control, and forward error correction are implemented in QUIC. 
In addition, QUIC incorporates cyptographic shields, such as IP spoofing protection and packet reordering \cite{cui2017innovating}. 
To keep the QUIC implementation lightweight and abstracted, we segregated the implementation into two parts: 
(i) the QUIC client and server only deal with UDP sockets and streams;
(ii) the agents deal with reliability and security.

\begin{figure}[t]
    \centering
    \includegraphics[width=1 \linewidth]{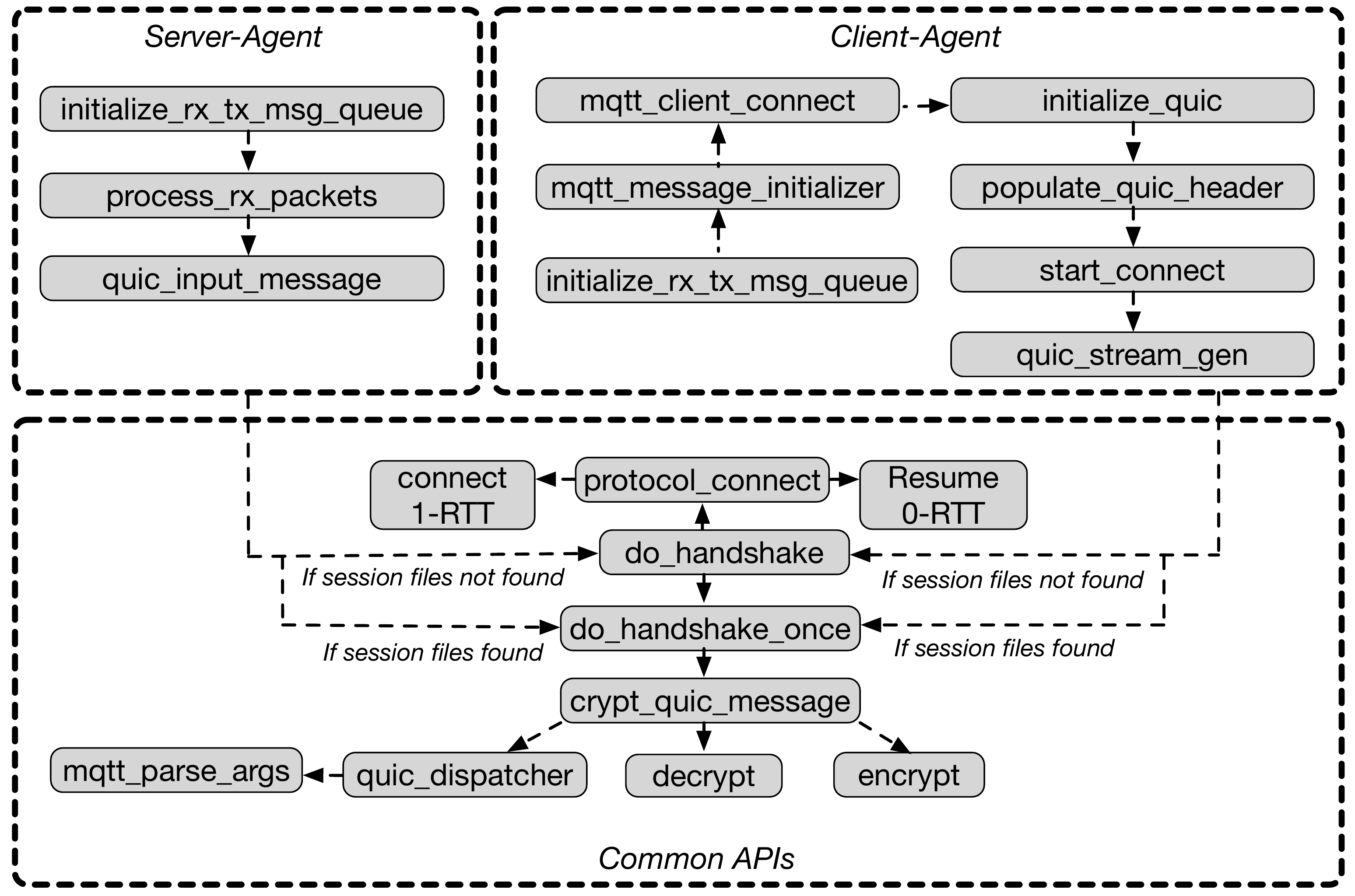}
    \captionsetup{font=scriptsize}
    \caption{Server-agent and client-agent are two entities between QUIC and MQTT. 
    This figure represents the packet processing flow by the server and client agents.}
    \label{fig:APIEngine}
\end{figure}

Figure \ref{fig:APIEngine} shows the high-level view of the server and client agent implementations.
The implemented entities are as follows:
(i) \textit{Server-Agent APIs and functions}: They handle server specific roles such as accepting incoming UDP connections, setting clients' state, and storing and forwarding packets to the subscribers based on topics. 
(ii) \textit{Client-Agent APIs and functions}: They perform client-specific tasks such as opening a UDP connection, and constructing QUIC header and streams. 
(iii) \textit{Common APIs and functions}: They are utilized by both the server and client. 
However, their tasks differ based on their roles.

After initializing the transmit and receive message queues by \texttt{initialize\_rx\_tx\_msg\_queue()}, the server and client agents process the messages differently. 
In the server-agent, the received message queue is fed to \texttt{quic\_input\_message()}. 
This API begins the handshake process to negotiate the session keys. 
However, if the client has contacted the server in the past, then \texttt{quic\_input\_message()} is directly available for the \texttt{crypt\_quic\_message()} to parse the incoming MQTT message. 
On the other hand, the client-agent first processes the MQTT message and then sends that message to the server by using \texttt{start\_connect()}. 
Based on the communication history between the server and client, the \texttt{start\_connect()} API either enters or skips the handshake process with the server. 
The common APIs handle the handshake control messages.
%
%
%
%


\begin{table*}[htp] 
\begin{threeparttable}
\begin{center}
\caption{Key Notations and Abbreviations}
\begin{tabular}{|c||l|}
\hline
\textbf{Variable}                    & \multicolumn{1}{c|}{\textbf{Description}}                                                                                                                                                                                                                                \\ \hline \hline

$\perp$                             & Represents a rejected message \\ \hline

$\lambda$                             & Protocol-associated security parameter used to derive the session keys\\ \hline

\textbf{$\kappa$}                    & Input key $\{0,1\}^\kappa \xrightarrow[]{\mathbb{\$}}\kappa$     \\ \hline                                                                                                                                                                                                                                                       
\textbf{$\mathcal{E}$}                  & Deterministic algorithm used in AEAD                                                                                                                                                                                                                \\ \hline
\textbf{$\mathcal{H}$}                & SHA-256 function          \\ \hline                                                                                              
$\sigma$                            & Signature used in digital signature scheme \\ \hline

\textbf{$(pk,sk)$}                     & Key pair representing public key and secret key                                                                                                                                                                                                           \\ \hline

\texttt{a}                     & Derived by $\{primes \ of \ size \ \lambda \} \xrightarrow[]{\mathbb{\$}} a$                                                                                                                                                                                                \\ \hline

\textbf{$action\_flag$}                & Flag used to determine if packet is subject to transmission or processing                                                                                                                                                                                                \\ \hline

\texttt{b}                     & Derived by $\{generators \ of \  \mathbb{Z}_a \} \xrightarrow[]{\mathbb{\$}} b$                                                                                                                                                                                                   \\ \hline

$C$                                     & Client (i.e., publisher or subscriber) \\ \hline

\textbf{\texttt{CHLO}}                        & Client hello message in QUIC, a.k.a., inchoate hello (\texttt{c\_i\_hello})                                                                                                                                                                                                  \\ \hline
\textbf{\texttt{cid}}                         & Connection identifier     \\ \hline                                                                                                                                                                                           $client\_info$                           & Struct used in storing client info such as \texttt{cid}, socket, etc.  \\ \hline               

\textbf{$client\_initial$}             & Client's initial state, which is the state immediately after receiving \texttt{REJ} message from server                                                                                                                                                                                                              \\ \hline

$\texttt{D}$                                    & Deterministic algorithm used in AEAD  \\ \hline

DH                                      & Diffie-Helman public values \\ \hline

\texttt{dup}                                  & \begin{tabular}[c]{@{}l@{}}Flag in MQTT (value 1 indicates the packet is a retransmission)
\end{tabular} \\ \hline


\texttt{HMAC}                                 & Key hash message authentication used in expansion of keys \cite{krawczyk1997hmac}                                                                                                                     \\ \hline
$ik$                                   & \begin{tabular}[c]{@{}l@{}}Initial key variable set in the initial phase of QUIC's connection establishment \end{tabular}                          \\ \hline

\texttt{init}                                   & \begin{tabular}[c]{@{}l@{}} bit initializer, \texttt{init} $\in \{0,1\}$\end{tabular}                          \\ \hline

\texttt{iv}                                   & \begin{tabular}[c]{@{}l@{}}Initialization vector\end{tabular}                                \\ \hline
$k$                                    & Key to derive data in \texttt{final\_data()} phase                                                                                                                                       \\ \hline
\texttt{Kg}                          & Key generation algorithm takes $\lambda$ as security parameter and generates ($pk$,$sk$) as key pair in QUIC                                    \\ \hline
\texttt{$k_{stk}$}           & Derived same as session key (for simplicity, in our implementation we treat \texttt{$k_{stk}$} as random string to replicate unpredictable input)

\\ \hline 
$M$                           & Message to be sent or received by client or server in QUIC                                                                                                                                                                                                              \\ \hline
$m$                           & The message part of \texttt{Msg} \\ \hline

\texttt{msgid}                       & Message ID for MQTT queuing                                                                                                                                                                                                                                              \\ \hline
\texttt{Msg}                       &\begin{tabular}[c]{@{}l@{}} Message space used in QUIC \\(\texttt{Msg} consists of bitstrings starting with a 1, while key
exchange messages to be encrypted start with a 0)\end{tabular}                                                                                                                                                                                                                                                \\ \hline

\texttt{nonc}                        & Nonce                                                                                                                                                                                                                                     \\ \hline
\texttt{pub}                         & Public DH Values                                                                                                                                                                                                   \\ \hline
\texttt{pk}                         & Public key                                                                                                                                                                                                  \\ \hline

\texttt{REJ}                         & Reject message (message from server after \texttt{CHLO} in QUIC)                                                                                                                                                                                                                  \\ \hline
\texttt{retained}                    & Flag used in MQTT to prevent loosing subscribed topics when a connection loss happens                                                                                                                                                                                                                             \\ \hline
RTT                         & Round-trip time       \\ \hline

$S$                         & Server \\ \hline

\texttt{scfg}                        & Variable (output by \texttt{scfg\_gen()}) to represent the global state of server in QUIC                                                                                                                                                                           \\ \hline

\texttt{scid}               & Server's \texttt{cid} \\ \hline

\texttt{SHLO}                        & Server Hello packet (\texttt{s\_hello}) in QUIC \\ \hline  

\texttt{sk}                         & Secret key                                                                                                                                                                                                   \\ \hline

\texttt{sqn}                         & Sequence number (used in signalling for every segment in QUIC)                                                                                                                                                                                        \\ \hline
$stk$                         & Source address token to guard IP-spoofing in QUIC                                                                                                                                                                           \\ \hline
$strike_{rng}$                 & Strike variable used by QUIC                                                                                                                                                                                                                                                                                                        \label{table:RefTable}                                                                                                                               \\ \hline
\end{tabular} 


\end{center} 
     
\end{threeparttable}
\end{table*}

\subsection{Definitions, Methods, and Assumptions} 
\label{definition} 
This section presents the primitives used to present and explain the implementations.
Table \ref{table:RefTable} shows the main acronyms and notations used in the rest of this paper.
We chose most of these notations based on the RFCs published relevant to this work.

Notation $\{0,1\}^* $ represents the set of all finite-length binary strings.
If $a$ and $b$ are two uniquely decodable strings, then ($a, b$) simply represents the concatenation of both. 
If $\kappa \in \mathbb{N}$, then $1^\kappa$ represents the $\kappa$ consecutive strings of 1 bits.
The notation $s \xrightarrow[]{\text{\$}}$ \texttt{S} represents the uniform random selection of $s$ from a finite set \texttt{S}. 
The set of integers [$1,.....n$] is represented by [$n$], where $n \in \mathbb{N}$.
We also assume that a public key infrastructure (PKI) is available. 
This means that public keys are user-identity bounded, valid, and publicly known. 
Therefore, certificates and their verification are excluded from the implementation.

A \textit{Digital Signature Scheme} (\texttt{SS}) with message space $Msg$ is used by the broker during a connection establishment to authenticate certain data.
The scheme is defined as follows,
\begin{equation}
\texttt{SS} = (\texttt{Kg}, \texttt{Sign}, \texttt{Ver})
\end{equation}
where $\texttt{Kg}$, $\texttt{Sign}$ and $\texttt{Ver}$ are the randomization key generation algorithm, signing algorithm, and verification algorithm, respectively. 
The input of $\texttt{Kg}$ is the security parameter $\lambda$ and its output is a public and secret key pair, as follows,
\begin{equation}
\texttt{Kg}(\lambda)\xrightarrow[]{\$}(pk,sk)
\end{equation}
The signing algorithm returns a signature,
\begin{equation} \label{eq:sign}
\texttt{Sign}(sk,m)\xrightarrow[]{\$}\sigma
\end{equation}
where $sk$ is the secret key and $m \in \texttt{Msg}$.
The verification algorithm is denoted as follows,
\begin{equation} \label{eq:ver}
\texttt{Ver}(pk, m, \sigma)\rightarrow p
\end{equation}
where $pk$ is the public key and $p \in \{0,1\}$. 
The output value $p$, which is a bit, shows whether the signature is valid or invalid.

The requirement for correctness of \texttt{SS} is \texttt{Ver}($pk$,$m$, \texttt{Sign}($sk$,$m$)) = 1 for every $m \in$ \texttt{Msg} and \texttt{Kg}($\lambda$). 
Correctness is defined by the requirement that the input of one party's $m_{send}$ be equal to the output of the other party's $m_{get}$. 

Secure channel implementation is based on an authenticated encryption with associative data schemes (AEAD) \cite{rogaway2002authenticated}. 
AEAD consists of two algorithms: $\mathcal{E}$ and $\texttt{D}$. 
First, $\mathcal{E}$ is a deterministic encryption algorithm defined as follows, 
\begin{equation} \label{eq:EP}
\mathcal{E}(k, nonc, m, H)\rightarrow c
\end{equation}
where $c$ is ciphertext, $nonc \in \{0,1\}^n$, message $m \in \{0,1\}^*$, $H \in \{0,1\}^*$ is an additional authenticated data, and key $k$ is defined as
\begin{equation} \label{eq:K}
\{0,1\}^\lambda\xrightarrow[]{\mathbb{\$}}\kappa
\end{equation}
Second, $\texttt{D}$ is a deterministic decryption algorithm defined as,
\begin{equation} \label{eq:D}
\texttt{D}(\kappa,nonc,H,\mathcal{E})\rightarrow pl \ or \perp
\end{equation}
where $pl$ is plaintext. 

The correctness requirement of AEAD is $\texttt{D}(\kappa$,\texttt{nonc},$H$,$\mathcal{E}$($\kappa$,\texttt{nonc},$H$,$m))$ \ $\equiv$ \ $m$ for all $\kappa \ \in \ \{0,1\}^\lambda$, \texttt{nonc} $\in \{0,1\}^n$, $H$ and $m$ $\in \{0,1\}^*$.

\begin{algorithm}[t]
\LinesNumbered
    \scriptsize
	\caption{\textbf{Server-agent APIs and functions}}
	\label{alg:server-agent}

	\footnotesize
	\SetInd{0.9em}{0.7em}

    \SetKwFunction{FMain}{main}
    \SetKwProg{Fn}{function}{}{}
    \Fn{\FMain{}}
    {
        /* Initialize tx and rx queues */\\
        \texttt{initialize\_rx\_tx\_msg\_queue();} \\
        $is\_client = false$; \\
        \While{true}{
            \uIf{event = process\_packet}{
                
                \uIf{packet must be transmitted}{
                     $action\_flag = send;$ \\
                }
                \Else {
                    $int \ sock = $\texttt{recvfrom();} \\
                    $client\_info.sock = sock;$ \\
                    \texttt{process\_rx\_packets();} \\
                }
            } \uElseIf{event = client\_disconnect $\parallel$ socket\_timeout}{
                disconnect $socket$; \\
                break; \\
            } 
        }    
    }

    \SetKwFunction{FMain}{process\_rx\_packets}
    \SetKwProg{Fn}{function}{}{}
    \Fn{\FMain{}}
    {

            \While{rx\_message\_queue} {
                    \texttt{quic\_input\_message();} \\
            }
        \KwRet\;
    }
	
    \SetKwFunction{FMain}{quic\_input\_message}
    \SetKwProg{Fn}{function}{}{}
    \Fn{\FMain{}}
    {
                \uIf{!(handshake\_completed)}{
                    \uIf{!(do\_handshake())}{
                        $client\_state$ = $client\_handshake\_failed$; \\
                        \KwRet $error$\;
                    } 
                    \Else{
                        $client\_state$ = $client\_initial;$  \\ 
                    }
        }
        \Else{
                $client\_state$ = $client\_post\_handshake$ ; \\
                \texttt{do\_handshake\_once()}; \\
        }
        \KwRet\;

    }

	
\end{algorithm}

\begin{algorithm}[!htb]
	\footnotesize
	\SetInd{0.9em}{0.7em}
	    \caption{\textbf{Client-agent APIs and functions}}
	\label{alg:Client-agent}
    \SetKwFunction{FMain}{main}
    \SetKwProg{Fn}{function}{}{}
    \Fn{\FMain{}}
    {
        \texttt{initialize\_tx\_rx\_msg\_queue()}; \\
        $is\_client = true$; \\
        \While{true}{
            \uIf{event = process\_packet}{
                     \uIf{packet must be transmitted}{
                         $int \ sock = $\texttt{create\_udp\_socket();} \\
                         $action\_flag = send$; \\
                         \texttt{mqtt\_message\_initializer();} \\
                     }
                \Else {
                    \texttt{insert}$(rx\_msg\_queue);$ \\
                    process \texttt{rx\_msg\_queue} for received packets; \\
                }
            } \uElseIf{event = client\_disconnect $\parallel$ socket\_timeout}{
                disconnect $socket$; \\
                $break;$ \\
            } 
        }    
        
    }
    \SetKwFunction{FMain}{mqtt\_message\_initializer}
    \SetKwProg{Fn}{function}{}{}
    \Fn{\FMain{}}
    {
        \uIf{(\texttt{!MQTTClient\_create()})}{
            \KwRet $error$\;    
        }  
        
        \uIf{(\texttt{!MQTTClient\_connectOptions\_initializer()})}{
            \KwRet $error$\; 
        }
        
        \uIf{(\texttt{!mqtt\_client\_connect()})}{
            \KwRet $error$\;
        }
        \KwRet\;
    }
    \SetKwFunction{FMain}{mqtt\_client\_connect}
    \SetKwProg{Fn}{function}{}{}
    \Fn{\FMain{}}
    {
        \uIf{$!(sanity(this\rightarrow msg))$}{
            \KwRet error\;  
        }   
        \texttt{initialize\_quic();} \\
        \KwRet\;
    }
    \SetKwFunction{FMain}{initialize\_quic}
    \SetKwProg{Fn}{function}{}{}
    \Fn{\FMain{}}
    {
        /* filling quic header */ \\
        \uIf{!$(client \rightarrow cid)$}{
              $client \rightarrow cid$ = \texttt{get\_cid();} /*Assign \texttt{cid} to client */\\
        }
        \texttt{populate\_quic\_header();} \\
        \texttt{start\_connect();} \\
        \KwRet\;
    }
    
     \SetKwFunction{FMain}{start\_connect}
    \SetKwProg{Fn}{function}{}{}
    \Fn{\FMain{}}
    {
        \texttt{quic\_stream\_gen();} \\
        \uIf{!$(handshake\_completed)$}{
            \texttt{do\_handshake();} \\
        } \Else{
            \texttt{do\_handshake\_once();} \\
        }
        \KwRet\;
    }
     \SetKwFunction{FMain}{quic\_stream\_gen}
    \SetKwProg{Fn}{function}{}{}
    \Fn{\FMain{}}
    {
        /* Check stream presence */ \\
        \uIf{$stream\_present$}{ 
            \uIf{$if\_stream\_closed$}{
                \texttt{clear\_entry();}  \\
                \KwRet $error$\;
            }
            \Else{
                /* Find stream */ \\
                $stream = $\texttt{find\_stream();} \\
            }
        } 
        \Else{
            /* Create a stream and associate it with the \texttt{cid}*/ \\
            $stream\rightarrow cid = \texttt{cid};$ \\
            $stream = $\texttt{create\_stream\_buf()}; \\
            
        }
          \KwRet; 
    }
\end{algorithm}

\begin{algorithm}[!htb]
\LinesNumbered
	
	\setcounter{AlgoLine}{0}
    \scriptsize
	\caption{\textbf{Common APIs and functions}}
	\label{alg:CommonAPIs}
	\footnotesize
	\SetInd{0.9em}{0.7em}
	\SetKwFunction{FMain}{do\_handshake}
	
    \SetKwProg{Fn}{function}{}{}
    \Fn{\FMain{}}
    {
        
        \uIf{!({protocol\_connect})}{
            \KwRet $error$\;
        }
        $handshake\_completed = true;$ \\
        \uIf{!(\texttt{do\_handshake\_once()})}{
            \KwRet $error$\;
        }

        \KwRet\;
    }
    
    \SetKwFunction{FMain}{do\_handshake\_once}
    \SetKwProg{Fn}{function}{}{}
    \Fn{\FMain{}}
    {
    
        \uIf{$action\_flag$ = $send$}{
            $flag$ = $encrypt$ ; \\
         }
         \Else{
            $flag$ = $decrypt$ ; \\ 
         }
         
        \uIf{\texttt{!(crypt\_quic\_message())}}{
                \KwRet $error$\;
        }

        \KwRet\;
    }
    
    \SetKwFunction{FMain}{crypt\_quic\_message}
    \SetKwProg{Fn}{function}{}{}
    \Fn{\FMain{}}
    {
            \uIf{flag =  decrypt}{
                 $this\rightarrow msg$ = \texttt{decrypt\_message}$(this\rightarrow msg);$ \\
                 \texttt{quic\_dispatcher();} \\
                 
            }
            \uElseIf{flag = encrypt}{
                $this\rightarrow msg$ = \texttt{encrypt\_message}$(this\rightarrow msg);$ \\
                \texttt{insert}$(this\rightarrow msg,\ tx\_msg\_queue);$ \\
            }
            
        \KwRet\;
    }
	
	\SetKwFunction{FMain}{protocol\_connect}
    \SetKwProg{Fn}{function}{}{}
    \Fn{\FMain{}}
    {
        \uIf{session files not found}{
            \texttt{connect()}; /* 1-RTT Implementation */\\ 
        } 
        \Else{
            \texttt{resume()}; /* 0-RTT Implementation */\\
        }
        \KwRet\;
        
    }

\SetKwFunction{FMain}{quic\_dispatcher}
    \SetKwProg{Fn}{function}{}{}
    \Fn{\FMain{}}
    {
        
        \uIf{$is\_client$}{
            /* If the client is calling the API */ \\
            process \texttt{rx\_msg\_queue} for mqtt application; \\
            \KwRet\;
        } 
        \Else{
            /* MQTT parsing for received messages */ \\
            \uIf{(\texttt{valid\_mqtt\_header($this\rightarrow msg$)})}{
                \texttt{mqtt\_parse\_args} $(this\rightarrow msg);$ \\
                \KwRet\;
            }
            \Else {
                \KwRet $error;$\\
            }
        }
        \KwRet\;
    }


\end{algorithm}

\begin{algorithm}[!htb]
    \LinesNumbered
	 \ContinuedFloat
	\caption{\textbf{Common APIs and functions (\textit{continued}) }}
	 \scriptsize
	 	\footnotesize
	\SetInd{0.9em}{0.7em} 
	
     \SetKwFunction{FMain}{encrypt\_message}
    \SetKwProg{Fn}{function}{}{m}
    \Fn{\FMain{}}
    {
        
        \uIf{!($handshake\_completed$)}{
            /* Initial session key is used */ \\
            \texttt{get\_iv(H, $ik$)}$\rightarrow iv;$ \\
            \uIf{$iv$ was used}{
                \KwRet $\perp$; \\
            } 
            \Else{
                /* For client */ \\
                \uIf{is\_client}{
                  \KwRet (\texttt{H}, $\mathcal{E}(ik_c, iv,$ \texttt{H}, $m$); \\  
                } 
                /* For server */ \\
                \Else{
                  \KwRet (\texttt{H}, $\mathcal{E}(ik_s, iv,$ \texttt{H}, $m$); \\ 
                }
            }
        }
        \Else{
            /* The stored established key is used */ \\
            \texttt{get}$\_iv$(\texttt{H}, $k)\rightarrow iv;$ \\
            \uIf{$iv$ is used}{
                \KwRet $\perp;$ \\ 
            }
            /* For client */ \\
            \uIf{is\_client}{
                \KwRet (\texttt{H}, $\mathcal{E}(k_c, iv,$ \texttt{H}, $m$); \\
            }
            /* For server */ \\
            \Else{
                  \KwRet (\texttt{H}, $\mathcal{E}(k_s, iv,$ \texttt{H}, $m$); \\ 
                }
        }
        
    }

	\SetKwFunction{FMain}{decrypt\_message}
    \SetKwProg{Fn}{function}{}{m}
    \Fn{\FMain{}}
    {
        /* Extracting ciphertext*/ \\
        $m\rightarrow ci$;\\
        \uIf{!($handshake\_completed$)}{
            /* initial data phase key is used */ \\
            \texttt{get\_iv(H, $ik$)}$\rightarrow iv;$ \\
                /* For client */ \\
                \uIf{is\_client}{
                    \uIf{\texttt{D}($ik_c$, $iv$, \texttt{H},         \texttt{ci}) $\neq$ $\perp$}{ 
                  
                        \KwRet $plain\_text$; \\
                    }
                    \Else{
                        \KwRet $\perp;$ \\
                    }
                }
                \Else{
                     /* For Server */ \\
                    \uIf{\texttt{D}($ik_s$, $iv$, \texttt{H},         \texttt{ci}) $\neq$ $\perp$}{ 
                  
                        \KwRet $plain\_text$; \\
                    }
                 }
              
        }
        \Else{
            /* Final data phase key is used */ \\
            \texttt{get\_iv(H, $k$)}$\rightarrow iv;$ \\
                /* For client */ \\
                \uIf{is\_client}{
                    \uIf{\texttt{D}($k_c$, $iv$, \texttt{H},         \texttt{ci}) $\neq$ $\perp$}{ 
                  
                        \KwRet $plain\_text$; \\
                    }
                    \Else{
                        \KwRet $\perp;$ \\
                    }
                }
                \Else{
                    /* For server */ \\
                    \uIf{\texttt{D}($k_s$, $iv$, \texttt{H},         \texttt{ci}) $\neq$ $\perp$}{ 
                  
                        \KwRet $plain\_text$; \\
                    }
                 }
        }
    }

\end{algorithm}	
	

\subsection{Server-agent}
Algorithm \ref{alg:server-agent} shows the implementation of these APIs and functions.
Our server-agent implementation is event-based. 
When an event $packet\_process$ occurs, the server-agent determines whether to transmit the packet or insert it into the receiving queue. 
Server-agent APIs and functions are described as follows:

\subsubsection{\texttt{main()}} If the packet is subject to transmission, then \texttt{action\_flag} is set, otherwise the packet will be processed by \texttt{process\_rx\_packets()}.
This function also handles the disconnect events. 
If the server receives a disconnect event, then particular clients will be disconnected. 
However, the server might receive a disconnect event for all the clients.
This happens, for example, when a reboot event occurs.

\subsubsection{\texttt{process\_rx\_packets()}} 
It processes the packets in the receive queue by \texttt{quic\_input\_message()}.

\subsubsection{\texttt{quic\_input\_message()}} 
This API first checks whether the handshake process with the client has been completed or not.
In order to determine this, it checks the $handshake\_completed$ flag. 
If it has been set, then it assumes that the connection is alive and handshake has been completed.
In this case, the API skips the handshake process and enters the \texttt{do\_handshake\_once()} API to set the flag for encryption or decryption. 
If the $handshake\_completed$ flag is not set, then the API runs the handshake process by executing \texttt{do\_handshake()}, where the client and server either follow the 1-RTT or 0-RTT implementation.
This API also sets the client's state according to its operation. 
For instance, if the client is contacting the server for the first time, then the client's state is set to \texttt{client\_initial}. 
If the handshake fails, then the client's state is set to \texttt{client\_handshake\_failed}. 
If the handshake has to be skipped, then it sets the client's state to \texttt{client\_post\_handshake}.

\subsection{Client-agent}
This section deals with the APIs and functions specifically used by the client-agent. 
Algorithm \ref{alg:Client-agent} shows the client agent implementation.

\subsubsection{\texttt{main()}} Similar to the server-agent, the \texttt{main()} function initializes both the transmit and receive queues by calling \texttt{initialize\_rx\_tx\_queue()}. 
The client-agent is also event-driven and processes both incoming and outgoing packets based on the \textit{packet\_process} event. 
The \texttt{main()} function checks $action\_flag$ to determine if the packet is meant to be sent or processed by \texttt{process\_rx\_packets()}.

\subsubsection{\texttt{mqtt\_message\_initializer()}} 
This API creates an MQTT client instance via \texttt{MQTTClient\_create()} and sets the client ID, the persistence parameter (\texttt{retained}), and the server IP address. 
When a subscriber connects to a broker, it creates subscriptions for all the topics it is interested in.
When a reboot or reconnect event occurs, the client needs to subscribe again. 
This is perfectly normal if the client has not created any persistent sessions.
The persistence parameter creates a persistent session if the client intends to regain all the subscribed topics after a reconnection or reboot event. 

Immediately after the creation of a client instance, a message is initialized by \texttt{MQTTClient\_connectOptions\_initializer()}. 
This API sets the \texttt{msgid}, \texttt{dup}, \texttt{retained}, \texttt{payload} and \texttt{version} fields of this packet.
\texttt{MQTTClient\_connectOptions\_initializer()} creates a directory to store all the topics and includes this in $client\_info$ to be easily retrieved later. 

\subsubsection{\texttt{mqtt\_client\_connect()}} 
At this point, all the message construction functionalities related to MQTT are completed, and finally the client instance is ready for QUIC related operations.
Sanity checking (e.g., header size, data length, etc.) is performed before entering the main QUIC API (i.e., \texttt{initialize\_quic()}).


\begin{algorithm}[!ht]
\LinesNumbered
	
	\setcounter{AlgoLine}{0}
    \scriptsize
	\caption{\textbf{1-RTT Implementation}}
	\label{alg:1-RTT}
	\footnotesize
	\SetInd{0.9em}{0.7em}
    \SetKwFunction{FMain}{connect}
    \SetKwProg{Fn}{function}{}{}
    \Fn{\FMain{}}
    {
    
        \uIf{\texttt{initial\_key\_exchange()}}{
            \uIf{\texttt{initial\_data()}}{
                \uIf{\texttt{key\_settlement()}}{
                    \uIf{\texttt{final\_data()}}{
                        \KwRet\ $success$; 
                    }
                }
            } 
        }
        \KwRet\ $error$;
    }

    \SetKwFunction{FMain}{initial\_key\_exchange}
    \SetKwProg{Fn}{function}{}{}
    \Fn{\FMain{}}
    {
        $message \ m$; \\
        /* For client */ \\
        \uIf{is\_client}{  
             \Switch{phase}{
                \Case{initial}{
                     /* $pk$ is public key */ \\
                    \texttt{c\_i\_hello($pk$)}$\rightarrow m_1$;  \\
                    \texttt{break}; \\
                }
                \Case{received\_reject}{
                    /* $m_2$ received packet from server (\texttt{REJ}) */ \\
                    \texttt{c\_hello($m_2$)}$\rightarrow m_3;$ \\
                    \texttt{break}; \\
                }
                \Case{complete\_connection\_establishment}{
                    /* $ik$ is initial key variable set during initial phase */ \\
                    \texttt{get\_i\_key\_c($m_3$)}$\rightarrow ik;$ \\
                    \texttt{break}; \\
                }
                
             }
        }
        /* For server */ \\
        \Else{
            \Switch{phase}{
                \Case{received\_chlo}{
                    \texttt{s\_reject($m_1$)} $\rightarrow m_2;$ \\ 
                     \texttt{break}; \\
                }
                \Case{received\_complete\_chlo}{
                     \texttt{get\_i\_key\_s($m_3$)} $\rightarrow ik;$ \\
                     \texttt{break}; \\
                }
            }
        }
        \KwRet\;
    }
    
     \SetKwFunction{FMain}{initial\_data}
    \SetKwProg{Fn}{function}{}{}
    \Fn{\FMain{}}
    {
        /* For client */ \\
        \uIf{is\_client}{
             
            for each $\alpha \in [i]$; \\
            $\alpha$ + 2 $\rightarrow$\texttt{sqn}$_c$; \\ 
            
            /* \texttt{sqn$_c$} = Client sequence number */ \\
            /* $M^\alpha_c$ = Client constructed message */ \\
            
            \texttt{pak($ik$, sqn$_c$, $M^\alpha_c$)} $\rightarrow m^\alpha_4;$ \\
            \texttt{process\_packets($ik,m_5$)}; \\
        }
        /* For server */ \\
        \Else{
            for each $\beta \in [j]$; \\
            
            $\beta$ + 1 $\rightarrow$ \texttt{sqn}$_s;$ \\

             /* \texttt{sqn$_s$} = Server sequence number*/ \\
             /* $M^\alpha_s$ = Server constructed message */ \\
            
            \texttt{pak($ik$, sqn$_s, M_s^\beta$)} $\rightarrow m;$ \\
            
            \texttt{process\_packets($ik, m_4$)}; \\
        }
        \KwRet\;
    }

\end{algorithm} 

\begin{algorithm}[!htb]
    \LinesNumbered
	 \ContinuedFloat
	\caption{\textbf{1-RTT Implementation (\textit{continued}) }}
	 \scriptsize
	 	\footnotesize
	\SetInd{0.9em}{0.7em} 
	\SetKwFunction{FMain}{key\_settlement}
    \SetKwProg{Fn}{function}{}{}
    \Fn{\FMain{}}
    {
        \uIf{is\_client}{
            
            \texttt{get\_key\_c($m_6$, sqn$_s$)} $\rightarrow k;$ \\
        }
        \Else{
            2 + $j$ $\rightarrow$ \texttt{sqn$_s$}; \\ 
            
            \texttt{s\_hello($m, ik$, sqn$_s$)} $\rightarrow m_6;$ \\
            
            \texttt{get\_key\_s($m_6)$} $\rightarrow k;$ \\
        }
        \KwRet\;
    }
	
 \SetKwFunction{FMain}{final\_data}
    \SetKwProg{Fn}{function}{}{}
    \Fn{\FMain{}}
    {
        \uIf{is\_client}{  
            for each $\alpha \in$ \{$i$ + 1,...,$u$\} \\
            $\alpha$ + 2 $\rightarrow$\texttt{sqn$_c$}; \\ 
            
            \texttt{pak($k$, sqn$_c$, $M_c^\alpha$)}$\rightarrow m_7^\alpha;$ \\

            $(m_7^{i+1}......m^u) \rightarrow m_7;$ \\
            \texttt{process\_packets}$(k,m_7)$; \\
        }
        \Else{
             for each $\beta \in$ \{$j$ + 1,...,$w$\} \\
             $\beta$ + 2 $\rightarrow$\texttt{sqn}$_s;$ \\
             
             \texttt{pak($k$, sqn$_s$, $M_s^\beta$)} $\rightarrow m_8^\beta;$ \\
    
             $(m_8^{j+1}......m^w) \rightarrow m_8;$ \\

             \texttt{process\_packets($k$, $m_8$)}; \\
        }
        \KwRet\;
    }

\end{algorithm}

\subsubsection{\texttt{initialize\_quic()}} 
This API first determines if the client has an assigned \texttt{cid}, and if not, then a new \texttt{cid} is generated by $\{0,1\}^{64}\xrightarrow[]{\$} \texttt{cid}$.
The QUIC header is initialized after \texttt{cid} assignment. 
The non-encrypted part of the QUIC header consists of \texttt{cid}, diversification nonce, packet number or sequence number, pointer to payload, size of the payload, size of the total packet, QUIC version, flags and encryption level. 
The encrypted part of the QUIC header is made of frames, where each frame has a stream ID and a final offset of a stream.
The final offset is calculated as a number of octets in a transmitted stream. 
Generally, it is the offset to the end of the data, marked as the \texttt{FIN} flag and carried in a stream frame. 
However, for the reset stream, it is carried in a \texttt{RST\_STREAM} frame \cite{iyengar2017quic}.
Function \texttt{populate\_quic\_header()} fills both the encrypted and non-encrypted parts of the QUIC header and enters the API \texttt{start\_connect()} to connect the client.

\subsubsection{\texttt{start\_connect()}} 
This API is the entering point to the common APIs (described in Section \ref{common_apis}). 
Its first task is to find an appropriate stream for the connection.
Based on the \texttt{handshake\_completed} flag, \texttt{start\_connect()} API determines if the client and server have completed the handshake. 
If the handshake has been completed, then it means that QUIC connection is alive and data can be encrypted or decrypted.
If not, then it enters the handshake process by executing the \texttt{do\_handshake()} API.

\subsubsection{\texttt{quic\_stream\_gen()}} 
The only purpose of this API is to detect and create streams. 
As mentioned earlier, QUIC is capable of multiplexing several streams into one socket.
This API first detects whether there is already an open stream for a particular client, and if not, then it creates a new stream by running \texttt{create\_stream\_buf()}.

\subsection{Common APIs and functions} 
\label{common_apis}
Several APIs and functions such as encryption, decryption and processing transmission packets are mandated on both the server and client sides. 
Although the packets handled by these functions are different in the client-agent and server-agent, the underlying mechanisms are almost similar. Algorithm \ref{alg:CommonAPIs} shows the implementation.

\subsubsection{\texttt{do\_handshake()}} 
This API is the starting point of the QUIC connection establishment and key exchange process. 
It behaves as an abstraction of the handshake process. 
First, it calls \texttt{protocol\_connect()} for protocol communication. 
If \texttt{protocol\_connect()} is executed successfully, then the \textit{handshake\_completed} flag is set. 
Based on this flag, the client determines whether to execute the handshake process or skip it. 
Lastly, this API calls the \texttt{do\_handshake\_once()} API to set the encrypt or decrypt flag for further processing.

\subsubsection{\texttt{do\_handshake\_once()}} 
The primary responsibility of this API is to set the encryption or decryption flags. 
This decision is made based on the type of the next operation, which is transmission or processing.
If \textit{action\_flag} is set, then the packet must be encrypted and sent.
Finally, it enters the \texttt{crypt\_quic\_message()} API.

\subsubsection{\texttt{crypt\_quic\_message()}} 
This API is the starting point in establishing a secure connection. 
It checks the $flag$ to determine if the packet should undergo decryption (\texttt{decrypt\_message()}) or encryption (\texttt{encrypt\_message()}).

\subsubsection{\texttt{protocol\_connect()}} 
\begin{figure}[t]
    \centering
    \includegraphics[width=0.9\linewidth]{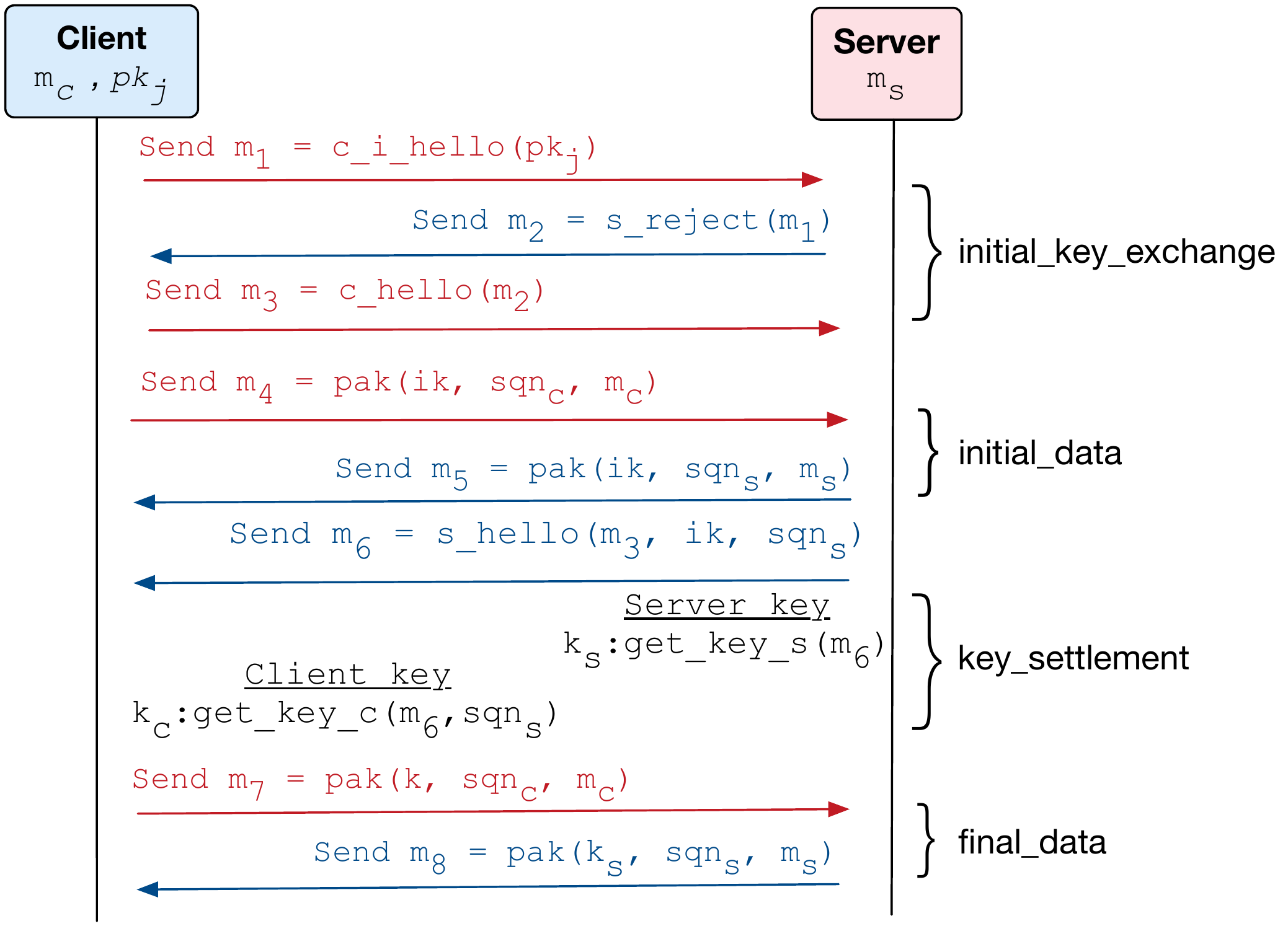}
    \captionsetup{font=scriptsize}
    \caption{{The four phases of the 1-RTT connection.}}
    \label{fig:1-RTT}
\end{figure}

\begin{figure}[t]
    \centering
    \includegraphics[width=0.98 \linewidth]{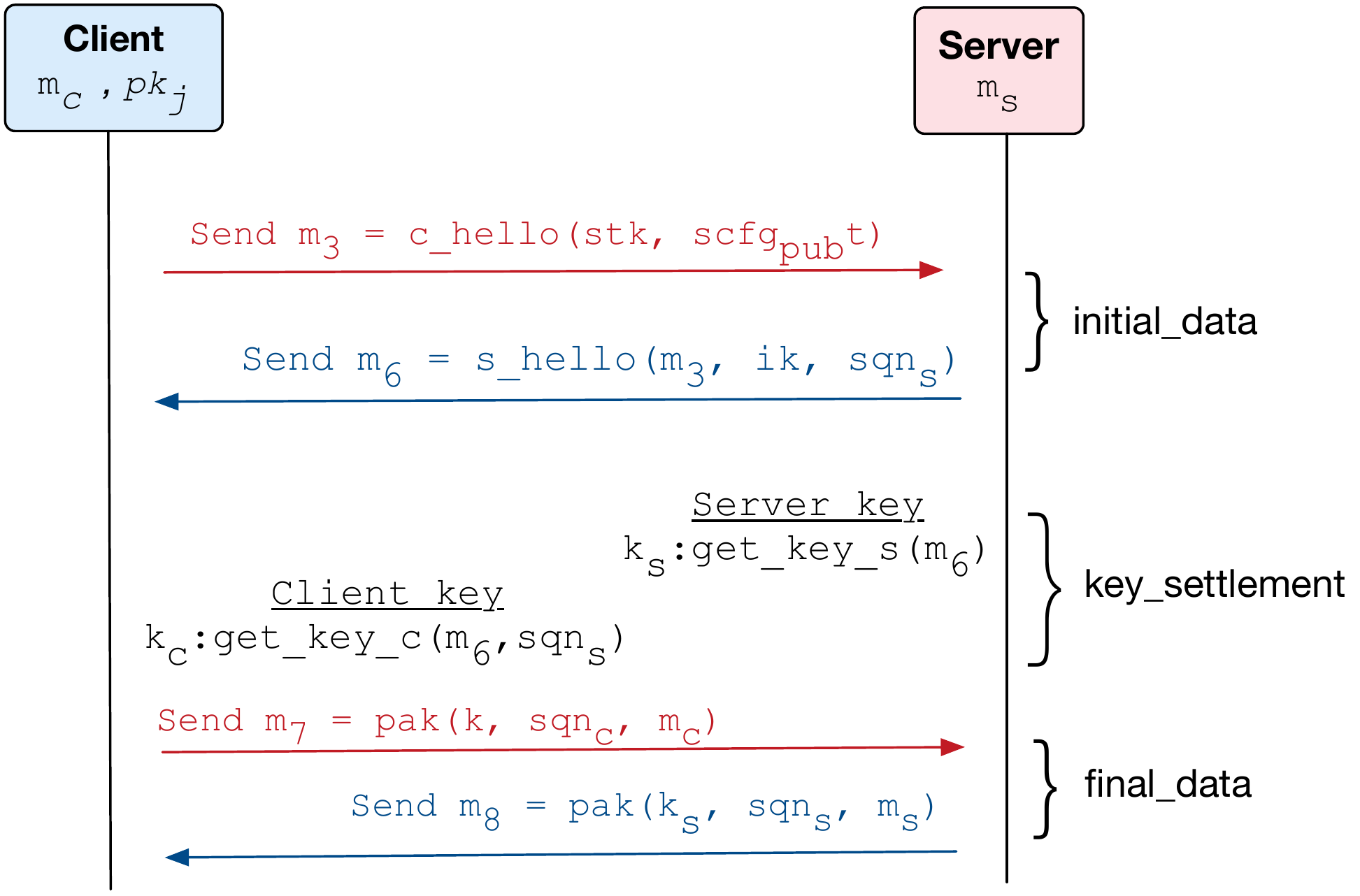}
    \captionsetup{font=scriptsize}
    \caption{{In 0-RTT scenario, client sends the first packet \texttt{c\_hello} with previous server global state scfg and strike. This step itself sends initial data to server.} }
    \label{fig:0-RTT}
\end{figure}

In this API, the client ($C$) checks the existence of a session file to determine whether it has communicated with the server ($S$) during the last $\tau_t$ seconds. 
If $C$ and $S$ are interacting for the first time, then the \texttt{connect()} API completes the 1-RTT scenario in four phases, as shown in Figure \ref{fig:1-RTT} and Algorithm \ref{alg:1-RTT}. 
If $C$ and $S$ have communicated before, then the \texttt{resume()} API follows the 0-RTT scenario shown in Figure \ref{fig:0-RTT} and Algorithm \ref{alg:0-RTT}.

\textbf{1-RTT Connection.} \label{1-rtt-text}
The 1-RTT implementation is divided into four phases. 
The first phase exchanges the initial keys to encrypt the handshake packets until the final key is set. 
The second phase starts exchanging encrypted initial data.
The third phase sets the final key.
Last, the fourth phase starts exchanging the final data.
We explain the details of these phases as follows.

%
%

\begin{algorithm}[t]
	\footnotesize
	\SetInd{0.9em}{0.7em}
	\caption{\textbf{Server Configuration State}}
	\label{alg:get_scfg}
    \SetKwFunction{FMain}{get\_scfg}
    \SetKwProg{Fn}{function}{}{}
    \Fn{\FMain{$sk, \tau_t, \lambda$}}
    {
    
        
        $\mathbb{Z}_{a-1} \xrightarrow[]{\mathbb{\$}} x_s;$ \\
        
        $b^{x_s} \xrightarrow{} y_s;$ \\
        
        $(b,a,y_s)\rightarrow$\texttt{pub}$_s;$ \\
        
        $x_s \rightarrow$\texttt{sec}$_s;$ \\
        
        $\tau_{t+1} \rightarrow$\texttt{expy}; \\
        
        $\mathcal{H}$(\texttt{pub}$_s$, \texttt{expy}) $\rightarrow$\texttt{scid}; /* $\mathcal{H}$ is a SHA-256 function */ \\ 
    
        \textit{"QUIC Server Config Signature"} $\rightarrow$  \texttt{str}; \\   
        
        \texttt{Sign($s_k$, (str, 0x00, \texttt{scid, pub}$_s$, \texttt{expy}))}$\rightarrow$\texttt{prof}; \\
        
        (\texttt{scid}, \texttt{pub$_s$}, \texttt{expy}) $\rightarrow$\texttt{scfg}$^t_{\texttt{pub}};$ \\
    
        return \texttt{scfg}; \\
    }
\end{algorithm}

\textbf{[1-RTT]: Phase 1.} 
\label{initialKeyExchange}
This phase is handled by \texttt{initial\_key\_exchange()} and consists of three messages, $m_1, m_2$ and $m_3$. 
The client $C$ runs \texttt{c\_i\_hello($pk$)}, which returns a packet $m_1$ with sequence number 1.
$m_1$ is an initial connection packet sent to $S$, containing a randomly generated \texttt{cid}.
In response to $m_1$, $S$ sends a \texttt{REJ} packet $m_2$, generated using \texttt{s\_reject($m_1$)}.
$m_2$ contains a source-address token $stk$ (similar to TLS session tickets \cite{salowey2008transport}), which is later used by $C$ to prove its identity to $S$ for the ongoing session and future sessions. 
This is performed by checking if the source IP address equals the IP address in $stk$.
Fundamentally, the $stk$ consists of an encryption block of $C$'s IP address and a timestamp. 
In order to generate $stk$, $S$ uses the same $\mathcal{E}$ deterministic algorithm (i.e., AEAD) with $k_{stk}$ (derived by $\{0,1\}^{128} \xrightarrow[]{\mathbb{\$}}k_{stk}$).
The initialization vector for $stk$ (i.e., \texttt{iv}$_{stk}$) is selected randomly and is used in \texttt{s\_reject}.
For simplicity, we implemented a validity range for $stk$, which is bounded by the time period during which it was either generated or set up. 
Another important parameter in $m_2$ is the $S'$s current state \texttt{scfg\_{pub}} (refer to Algorithm \ref{alg:get_scfg}). 
It contains $S$'s DH values with an expiration date and a signature \texttt{prof}. 
This signature is signed by \texttt{SS} over all the public values under the $S'$s secret key $sk$.  


Upon receiving $m_2$, the client $C$ checks \texttt{scfg}$_{pub}^t$ for its authenticity and expiration. 
The algorithms for this purpose can be found in \cite{rogaway2002authenticated}. 
As we mentioned in Section \ref{definition}, our implementation assumes that a PKI is in place. 
After possessing the public key of $S$, the client generates a \texttt{nonc} and DH values by running \texttt{c\_hello($m_2$)}, and sends them to the server in message $m_3$.
At this point, both $C$ and $S$ derive the initial key material $ik$, using \texttt{get\_i\_key\_c($m_3$)} and \texttt{get\_i\_key\_s($m_3$)},  respectively. 
The server keeps track of the used \texttt{nonc} values in order to make sure that it does not process the same connection twice.
This mechanism is referred as \textit{strike-register} or \texttt{strike}. 
The timestamp is included in the \texttt{nonc} by the client.
The server only maintains the state of a connection for a limited duration of time.
Any connection request from a client is rejected by the server if its \texttt{nonc} is already included in its \texttt{strike} or contains a timestamp that is outside the permitted range \texttt{strike}$_{rng}$.
The initial key $ik$ = ($ik_c$, $ik_s$, \texttt{iv}) is made of two parts: two 128-bit application keys ($ik_c, ik_s$) and two 4-byte initialization vector prefixes \texttt{iv} = (\texttt{iv$_c$}, \texttt{iv$_s$}). 
The client uses $ik_s$ and \texttt{iv$_s$} to encrypt the data and send it to $S$. 
On the other hand $ik_c$ and \texttt{iv$_c$} assist in decryption and encryption.
This phase happens only once during the time period $\tau_t$ until \texttt{scfg}$_{pub}^t$ and $stk$ are not expired.

\begin{algorithm}[!t]
    \LinesNumbered
	\setcounter{AlgoLine}{0}
    \scriptsize

    \caption{\textbf{QUIC messages exchange APIs and functions}}
	\label{alg:initial_key_exchange}
    
	\footnotesize
	\SetInd{0.9em}{0.7em}
        
    \SetKwFunction{FMain}{c\_i\_hello}
    \SetKwProg{Fn}{function}{}{}
    \Fn{\FMain{pk}}
    {
        $\{0,1\}^{64}\xrightarrow[]{\mathbb{\$}}$\texttt{cid}; \\
    
        \KwRet\ \{\texttt{IP}$_c$, \texttt{IP}$_s$, \texttt{port}$_c$, \texttt{port}$_s$, \texttt{cid}, 1\};
    }

    \SetKwFunction{FMain}{s\_reject}
    \SetKwProg{Fn}{function}{}{}
    \Fn{\FMain{m}}
    {
        $\{0,1\}^{96}\xrightarrow[]{\mathbb{\$}}$\texttt{iv}$_{stk}$; \\   
    
        (\texttt{iv}$_{stk}$, $\mathcal{E}(k_{stk}$, \texttt{iv}$_{stk}$, $\epsilon$, 0) $\parallel$ (\texttt{IP}$_c$, \texttt{current\_time}$_s$))$\rightarrow$ \texttt{stk}; \\ 
        
       \KwRet\ \{\texttt{IP$_s$,IP$_c$,port$_s$,port$_c$,\texttt{cid},1,\texttt{scfg}$_{pub}^t$, \texttt{prof}, \texttt{stk}\}}; \\
       /* \texttt{prof} is generated by \texttt{get\_scfg} */
    }
    
    \SetKwFunction{FMain}{c\_hello}
    \SetKwProg{Fn}{function}{}{}
    \Fn{\FMain{m}}
    {
       \uIf{expy $\leq$ \ $\tau_t$}{
            \KwRet\; 
       }
       $str = "QUIC \ server \ config \ signature"$; \\
       \uIf{\texttt{Ver$(pk, ($\texttt{str}$, 0X00$,\texttt{scid, pub$_s$, expy}), \texttt{prof})) $\neq 1; $}}{
            \KwRet\;
       }
       

       $\{0,1\}^{160} \xrightarrow[]{\mathbb{\$}} r$; \\
       $(current\_time_c, r)\rightarrow \texttt{nonc};$ \\
      
       $\mathbb{Z}_{a-1}\xrightarrow[]{\mathbb{\$}} x_c, \ b^{x_c} \rightarrow y_c,  \ (b,a,y_c) \rightarrow pub_c;$ \\
       $\texttt{(IP$_c$, IP$_s$, port$_c$, port$_s$)} \rightarrow $ \texttt{pkt\_info}; \\
       
       \KwRet\ (\texttt{pkt\_info}, \texttt{cid, 2, stk, scid, nonc, pub$_c$}); 
    }
    
    \SetKwFunction{FMain}{get\_i\_key\_c}
    \SetKwProg{Fn}{function}{}{}
    \Fn{\FMain{m}}
    {
       $y_s^{x_c} \xrightarrow[]{\mathbb{\$}}$\texttt{ipm}; \\
       \KwRet\ \texttt{extract\_expand(ipm, nonc, cid, $m$, 40, 1)} \\ 
    }
    
    \SetKwFunction{FMain}{get\_i\_key\_s}
    \SetKwProg{Fn}{function}{}{}
    \Fn{\FMain{m}}
    {
        \texttt{stk}$\rightarrow$(\texttt{iv$_{stk}$}, \texttt{tk}); \\
        
        $\texttt{D}(k_{stk}, \texttt{iv}_{stk}, \epsilon, tk) \rightarrow$ \texttt{d}; \\
       
       \uIf{(d = $\perp \ \parallel (\ first \ 4 \ bytes \ of \ d \neq 0)$}{
            \uIf{$(\ first \ 4 \ bytes \ of \ d) \ \neq  \ \texttt{IP}_c$)}{
                \KwRet; 
            }
            \KwRet;  
        }    
       
       \BlankLine
       
       \uIf{last 4 bytes corresponds to outside \texttt{strike}$_{rng}$}{
            \KwRet\;
       }
       
       \uIf{ $r \in \texttt{strike} \parallel \tau_t \notin \texttt{strike}_{rng}$}{
           \KwRet\; 
       }
       
       \uIf{\texttt{scid} is unknown}{
            \KwRet\; 
       }
       
       \uIf{\texttt{scid} corresponds to expired \texttt{scfg}$_{pub}^{t^`}$}{
            /* where $t^`$ \ \textless \ $t$ */ \\
            \KwRet\;
       }
       
       \uIf{$b, a \in \texttt{pub}_c \neq b, a \in \texttt{pub}_s$}{
            \KwRet\;
       }
       
       \BlankLine 
       
        $ y_c^{x_s} \rightarrow$\texttt{ipm}; \\    
       
       \KwRet\ \texttt{extract\_expand(\texttt{ipm,nonc,cid,m,}40,1)}; \\
    }

    \SetKwFunction{FMain}{extract\_expand}
    \SetKwProg{Fn}{function}{}{}
    \Fn{\FMain{ipm, nonc, cid, m, l, init}}
    {
         \texttt{HMAC(nonc,ipm)$\rightarrow$ ms}; \\
        \uIf{\texttt{init} = 1}{
            $"QUIC \ key  \ expansion" \rightarrow str;$ \\
        }
        \Else{
            $"QUIC \ forward \ secure \ key  \ expansion" \rightarrow str;$ \\
        }
        
        ($str, 0X00,$ \texttt{cid}, $m$, \texttt{scfg}$_{pub}^t$) $\rightarrow$\texttt{info}; \\
        
        \KwRet\ first $l$ bytes (octets) of \texttt{T = (T(1), T(2),...)},\\ if all $i \in \mathbb{N}$, 
        \texttt{T($i$) = HMAC(ms, (T($i$-1), info, 0x0$i$))} and \texttt{T(0) = $\epsilon$}
    }    
\end{algorithm}

\textbf{[1-RTT]: Phase 2.}
This phase is handled by \texttt{initial\_data()} and consists of two messages, $m_4$ and $m_5$.
The client $C$ and server $S$ exchange the initial data message $M_c$ and $M_s$, which are encrypted and authenticated with $ik$ in function \texttt{pak($ik$,sqn$_c$, $M_\alpha^i$)} for every $\alpha \in [i]$ and \texttt{pak($ik$, sqn$_s$, $M_\beta^i$)} for every $\beta \in [j]$, respectively. 
Here, \texttt{sqn$_c$} and \texttt{sqn$_s$} represent the sequence number of packets sent by $C$ and $S$, respectively. 
$i$ and $j$ represent the maximum number of message blocks that $C$ and $S$ can exchange prior to the first phase. 
The initialization vector \texttt{iv} is generated based on the server or client role. 
When $S$ sends a packet, \texttt{get\_iv()} outputs \texttt{iv} by concatenating \texttt{iv}$_c$ and \texttt{sqn}$_s$. 
When $C$ sends a packet, \texttt{get\_iv()} generates \texttt{iv} by concatenating \texttt{iv}$_s$ and \texttt{sqn}$_c$.
The total length of each \texttt{iv} is 12 bytes since both the server and client initialization vector prefixes (i.e., \texttt{iv}$_c$ and \texttt{iv}$_s$) are 4 bytes in length and sequence numbers (i.e., \texttt{sqn}$_s$ and \texttt{sqn}$_c$) are 8 bytes in length. 
When $C$ receives packets from $S$, it uses the \texttt{process\_packets()} function to decrypt those packets to extract their payloads and concatenates them based on their sequence number. 
The server $S$ performs a similar mechanism for the packets received from $C$.

\begin{algorithm}[!htb]
     \LinesNumbered
	 \ContinuedFloat
	\caption{\textbf{QUIC messages exchange APIs and functions (\textit{continued})}}
	 \scriptsize
	\label{alg:initial_data}
	\footnotesize
	\SetInd{0.9em}{0.7em}
      
   \SetKwFunction{FMain}{get\_iv}
    \SetKwProg{Fn}{function}{}{}
    \Fn{\FMain{$H,\kappa$}}
    {
    
        /* $\kappa_c = $ \texttt{Client key},  \\
         * $\kappa_s = $ \texttt{Server key} \\
         * $iv_c = $ \texttt{Client initialization vector}, \\
         * $iv_s = $ \texttt{Server initialization vector} */  \\
         
         $\kappa \rightarrow (\kappa_c, \ \kappa_s, \ iv_c, \ iv_s);$ \\
         
        \uIf{is\_client}{
            $c\rightarrow src, s\rightarrow dst;$ \\
        }
        \Else{
            $s \rightarrow src$, $c \rightarrow dst;$ \\
           
        }
        /* \texttt{sqn} is packet sequence number */ \\
        $H \rightarrow$\texttt{(cid, sqn)}; \\
        \KwRet\ ($iv_{dst}$, \texttt{sqn});
    }
    
     \SetKwFunction{FMain}{pak}
    \SetKwProg{Fn}{function}{}{}
    \Fn{\FMain{$k$, \texttt{sqn}, $m$}}
    {
        $\kappa \rightarrow (k_c,k_s$,\texttt{iv}$_c$, \texttt{iv}$_s$); \\
        \uIf{is\_client}{
            $c \rightarrow src$ and $s \rightarrow dst$; \\
        }
        \Else{
            $s \rightarrow src$ and $c \rightarrow dst$; \\
        }
        
        (\texttt{IP}$_{src}$, \texttt{IP}$_{dst}$, \texttt{port}$_{src}$, \texttt{port}$_{dst}$) $\rightarrow$ \texttt{pkt\_info}; \\ 
        
        \texttt{(cid, sqn)}$\rightarrow H;$ \\
        $\texttt{get\_iv}(H,\kappa) \rightarrow$ \texttt{iv}; \\
        
        \KwRet\ (pkt\_info, $\mathcal{E}$($k_{dst}$,\texttt{iv},$H$, (1 $\parallel m$)));  
    }
    
     \SetKwFunction{FMain}{process\_packets}
    \SetKwProg{Fn}{function}{}{}
    \Fn{\FMain{$\kappa,p_1, p_2....p_v$}}
    {
        $\kappa \rightarrow (k_c,k_s,\texttt{iv}_c,\texttt{iv}_s);$ \\
        \uIf{is\_client}{
            $c \rightarrow src$ and $s \rightarrow dst;$ \\
        }
        \Else{
            $s \rightarrow src$ and $c \rightarrow dst;$ \\
        }
        
        for each $ \gamma \in [v] :$ \\
        \hspace{0.35cm}$p_\gamma \rightarrow (H_\gamma, c_\gamma);$ \\ 
        \hspace{0.35cm}$\texttt{get\_iv}(H_\gamma, \kappa) \rightarrow \texttt{iv}_\gamma;$ \\
        \hspace{0.35cm}$\texttt{D}(k_{src},iv_\gamma,H_\gamma,c_\gamma) \rightarrow m\gamma;$ \\
        \hspace{0.35cm}\uIf{$m_\gamma \notin$ \texttt{Msg}}{
           \KwRet\
        }   
        
    \KwRet\ ($m_1$, $m_2$, .... $m_v$);

    }
    \SetKwFunction{FMain}{s\_hello}
    \SetKwProg{Fn}{function}{}{}
    \Fn{\FMain{$m_3$, $ik$, \texttt{sqn}}}
    {
        $ik \rightarrow (ik_c, ik_s, \texttt{iv}_c, \texttt{iv}_s);$ \\
        
        $\mathbb{Z}_{a-1} \xrightarrow[]{\mathbb{\$}} \widetilde{x_s}, \ b^{\widetilde{x}_s} \rightarrow \widetilde{y}_s, \ (b,a,\widetilde{y}_s) \rightarrow \texttt{p}\widetilde{\texttt{u}}\texttt{b}_s;$ \\  
    
        \texttt{(cid, sqn)}$\rightarrow H;$ \\
    
        $\mathcal{E}(ik_c, (\texttt{iv}_c, sqn), H, (0 \parallel (\texttt{p}\widetilde{\texttt{u}}\texttt{b}_s, \texttt{stk})))\rightarrow e$; \\

	    \KwRet\ (\texttt{ip}$_s$,\texttt{ip}$_c$,\texttt{port}$_s$,\texttt{port}$_c$, $H$, $e$); \\  
	}
	
	\SetKwFunction{FMain}{get\_key\_s}
    \SetKwProg{Fn}{function}{}{}
    \Fn{\FMain{$m$}}
    {
        $y_c^{\widetilde{\texttt{x}}_s} \rightarrow$\texttt{pms}; \\
	    \KwRet\ \texttt{extract\_expand(pms, nonc, cid, $m$, 40, 0)};
	}
	
	\SetKwFunction{FMain}{get\_key\_c}
    \SetKwProg{Fn}{function}{}{}
    \Fn{\FMain{$m$}}
    {
        $m \rightarrow$\texttt{(IP$_s$,IP$_c$,port$_s$, port$_c$, cid, sqn, $e$)}; \\
        \uIf{\texttt{$\texttt{D}$($ik_c$,(iv$_c$, sqn),(cid,sqn),$e$) = $\perp$}}{
            \KwRet\ 
        }
        \BlankLine
        \uIf{\texttt{first \ bit \ of \ the \ message $\neq$ 0}}{
            \KwRet\ 
        }
        \BlankLine
        ${\widetilde{y}}^{x_c}_s \rightarrow$ \texttt{pms}; \\
         \KwRet\ \texttt{extract\_expand(pms, nonc, cid, $m$, 40, 0)};
        
    }
 
\end{algorithm}

\begin{algorithm}[!htb]
	\footnotesize
	\SetInd{0.9em}{0.7em}
    \caption{\textbf{0-RTT Implementation}}
    \label{alg:0-RTT}
    \SetKwFunction{FMain}{resume}
    \SetKwProg{Fn}{function}{}{}
    \Fn{\FMain{}}
    {
        \uIf{(\texttt{pub}$_c$)}{
            \uIf{\texttt{stk}}{
                \uIf{\texttt{scid}}{
                    \KwRet\ \texttt{c\_hello(\texttt{stk}, \texttt{scfg}$_{pub}^t$);}  
                }    
            }
           
        }
         /* Jumping back on 1-RTT */ \\
        \KwRet\ \texttt{connect()}; 
    
	}

    \SetKwFunction{FMain}{c\_hello}
    \SetKwProg{Fn}{function}{}{}
    \Fn{\FMain{\texttt{stk}, \texttt{scfg}$_{pub}^t$}}
    {
        $\{0,1\}^{64}\xrightarrow[]{\mathbb{\$}}$\texttt{cid}; \\
    
        $\{0,1\}^{160} \xrightarrow[]{\mathbb{\$}} r$, (\texttt{current\_time$_c$, $r$}) $\rightarrow$ \texttt{nonc}; \\  
      
        $\mathbb{Z}_{a-1} \xrightarrow[]{\mathbb{\$}}x_c$, $b^{x_c} \rightarrow y_c$, $(b,a,y_c)\rightarrow$ \texttt{pub}$_c$; \\
        
        (\texttt{IP}$_c$, \texttt{IP}$_s$,port$_c$,port$_s$)$\rightarrow$\texttt{pkt\_info}; \\
        
        \KwRet\ (pkt\_info, \texttt{cid}, 1, \texttt{stk}, \texttt{scid}, \texttt{nonc}, \texttt{pub}$_c$)
	}
\end{algorithm}

\textbf{[1-RTT]: Phase 3.}
This phase is handled by \texttt{key\_settlement()} and involves message $m_6$. 
The server $S$ produces new DH values (authenticated and encrypted via AEAD with $ik$) and transmits them to the client using \texttt{s\_hello($m_3$,$ik$, sqn)}. 
The client verifies the server's new DH public values with the help of $ik$. 
At this point, the server and client both derive the session key by \texttt{get\_key\_s($m_6$)} and \texttt{get\_key\_c($m_6$)} and use \texttt{extract\_expand()} for key expansion, as defined in Algorithm \ref{alg:initial_key_exchange}.

\textbf{[1-RTT]: Phase 4.} 
\label{finalData}
This phase is handled by \texttt{final\_data()} and consists of two messages, $m_7$ and $m_8$. 
Instead of initial key $ik$, the established key $k$ is used to encrypt and authenticate the remaining data (\texttt{Msg}) by both $C$ and $S$.
Similar to $ik$, $k$ is derived from $k_c$, $k_s$, and \texttt{iv}, and consists of two parts: the two 128-bit application keys ($k_c$, $k_s$) and the two 4-bytes initialization vector prefixes \texttt{iv} = (\texttt{iv}$_c$,\texttt{iv}$_s$). 
In order to encrypt the data, $C$ uses $k_s$ and \texttt{iv}$_s$ before sending to $S$. 
For decryption, $C$ uses \texttt{iv}$_c$ and \texttt{k}$_c$ to decrypt the data received from $S$.

\textbf{0-RTT Connection.} 
\label{0-rtt-text}
Another scenario of connection establishment is 0-RTT, as Figure \ref{fig:0-RTT} shows. 
If $C$ has already established a connection with $S$ in the past $\tau_t$ seconds, then $C$ skips sending \texttt{c\_i\_hello()}, and initiates another connection request to the server by sending a \texttt{c\_hello()} packet. 
This packet contains the existing values of \texttt{stk}, \texttt{scid}, \texttt{cid}, \texttt{nonc}, and \texttt{pub}$_c$.
It is important to note that \texttt{pub}$_c$ requires new DH ephemeral public values. 
After receiving \texttt{c\_hello}, $S$ verifies if the \texttt{nonc} is fresh. 
This is performed against strike-register, provided that \texttt{stk} is valid and \texttt{scid} is not unknown or expired. 
If these verification steps do not succeed, then $S$ returns to the 1-RTT process by generating and sending out a \texttt{s\_reject} message (Algorithm \ref{alg:initial_key_exchange}). 
If these verification steps succeed, then the rest of the protocol remains the same. 
Algorithm \ref{alg:0-RTT} shows the implementation for 0-RTT.

\subsubsection{\texttt{quic\_dispatcher()}}
This API is the starting point where MQTT-related parsing starts. 
After the QUIC header is stripped off and decrypted, the packet is delivered to MQTT. 
Here, the broker finds the topic and delivers the message to all the subscribers that have subscribed to that topic.

\subsection{Code Reduction}
QUIC is a part of the Chromium Projects \cite{QUIC_Proto}. 
These projects have a heavy code size because they include several features such as the Chrome browser, SPDY protocol, Chromecast, Native Client, and QUIC with their entwined implementations. 
In order to reduce the code size, we have removed all these features except those that are essential to the functionality of QUIC. 
Furthermore, to reduce its overhead, most of the logging features have been removed or disabled.
We have also eliminated \textit{alarm\_factory}, which generates platform-specific alarms. 
The \textit{host resolver} has been removed as well.
As most clients have an in-built DNS pre-fetching \cite{krishnan2010dns} feature, they can resolve DNS queries without having to communicate with a server.
Since MQTT routes data flows based on topics instead of URLs, this feature is not required. 
We have also removed the backend proxy-related code.
Backend proxy is used when the proxy server is behind the firewall and load balances the requests from clients to the servers. 
Since MQTT is based on the publish/subscribe model, it is futile to involve any proxy or load balancing. 
Finally, we removed all the unit-testing code.

\section{Evaluation} 
\label{result_final}
This section presents the performance evaluation of MQTTw/QUIC versus MQTTw/TCP.
%
%

%
%
In order to measure the impact of the physical layer and distance on performance, three different testbeds are used: \textit{wired}, \textit{wireless}, and \textit{long-distance}.

\begin{itemize} [itemsep=0pt, topsep=0pt, leftmargin=*]
\item [--] \textbf{Wired.} All the nodes are connected to a Netgear 1Gbps L2-learning switch.
\item [--]  \textbf{Wireless.} The nodes communicate through an 802.11n network.
The access point used is a Linksys AC1200 and operates on channel 6 of the 2.4GHz band.
In addition, we placed two other similar access points 3 meters away form the main access point.
In order to introduce latency and packet drops, these two access points operate on the same channel and therefore interfere with the main access point.
Each interfering access point continuously exchanges a 10Mbps UDP flow with a nearby user.
\item [--] \textbf{Long Distance.} Our objective in using the long-distance routed network is to introduce longer and unpredictable end-to-end delays.
The subscriber and publisher are placed in Santa Clara, California, and the broker is situated across the country in Washington D.C.     
\end{itemize}

\begin{figure}[t]
    \centering
    \includegraphics[width=0.96\linewidth]{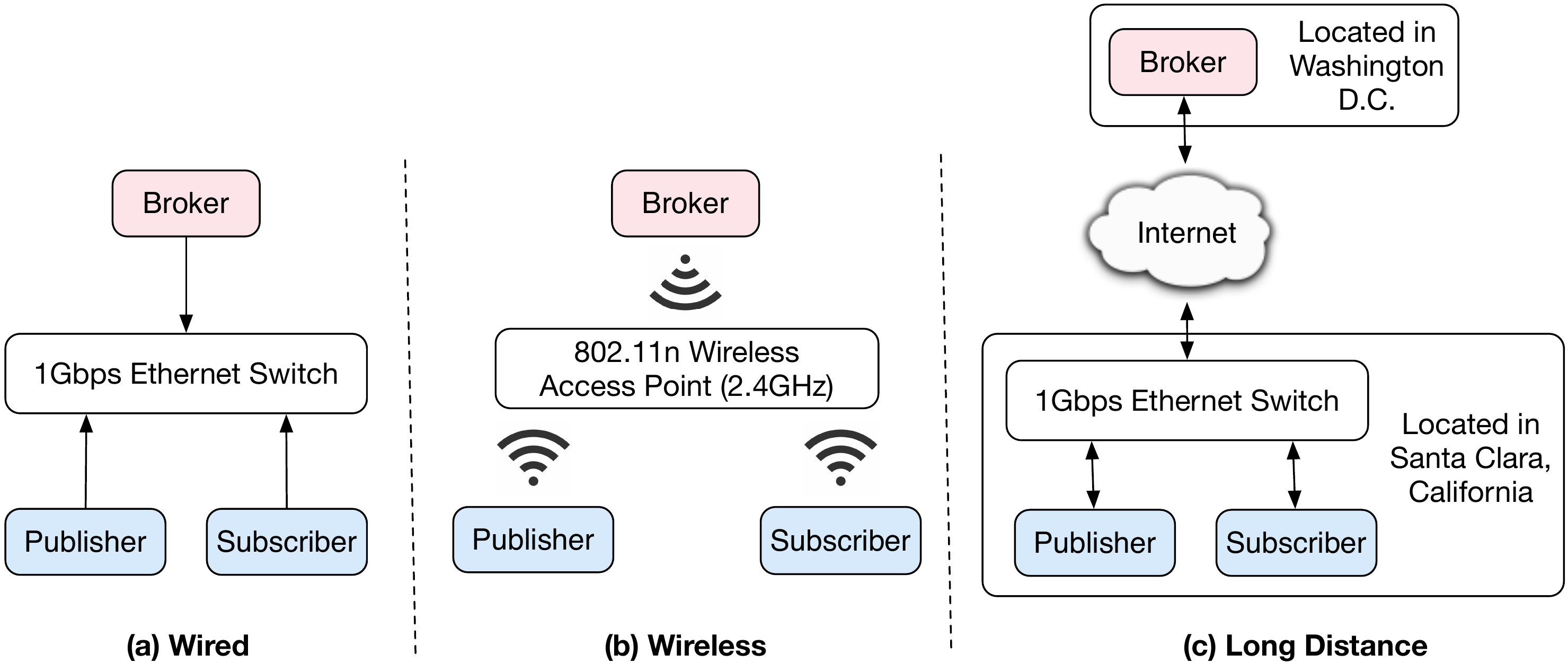}
    \captionsetup{font=scriptsize}
    \caption{The three testbeds used for the performance evaluations.}
    \label{fig:ScenariosTopo}
\end{figure}

Raspberry Pi 3 model B is the device used as the subscriber, publisher, and broker.
These devices run Raspbian Stretch as their operating system. 
As mentioned in the previous section, the MQTT implementation is based on Eclipse Paho and Mosquitto.
In addition, the TLS version is 1.2, the socket timeout for TCP and QUIC is 30 seconds, and the keep-alive mechanism of MQTT has been disabled.
Figure \ref{fig:ScenariosTopo} shows these testbeds.

When presenting the results, each point is the median of values obtained in 10 experiments, where each experiment includes 10 iterations.
For example, when measuring the overhead of connection re-establishment, the subscriber and publisher are connected to the broker 10 times, and the result is counted as one experiment.
The iterations of an experiment are run consecutively, and the minimum time interval between the experiments is 5 minutes.

The rest of this section studies the performance of MQTTw/QUIC versus MQTTw/TCP in terms of the overhead of connection establishment, head-of-line blocking, half-open connections, resource utilization, and connection migration.

\subsection{Overhead of Connection Establishment}
\label{ConserveResources}
\begin{figure}[!t]
    \centering
    \includegraphics[width=0.99\linewidth]{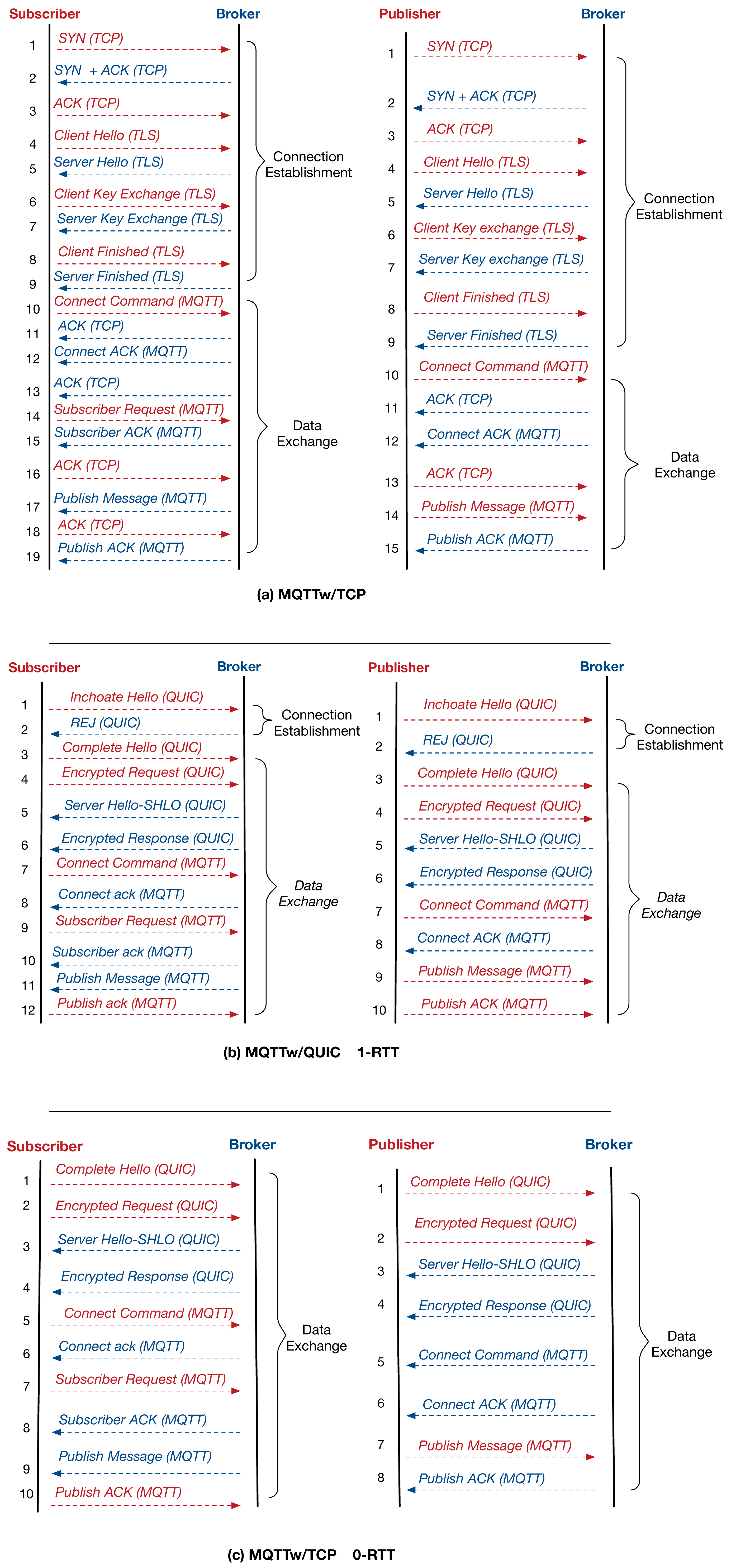}
    \captionsetup{font=scriptsize}
    \caption{The connection establishment process using (a) MQTTw/TCP, (b) MQTTw/QUIC 1-RTT, and (c) MQTTw/QIUC 0-RTT.}
    \label{fig:conn_pkt_exch}
\end{figure}

\begin{figure}[t]
    \centering
    \includegraphics[width=0.8\linewidth]{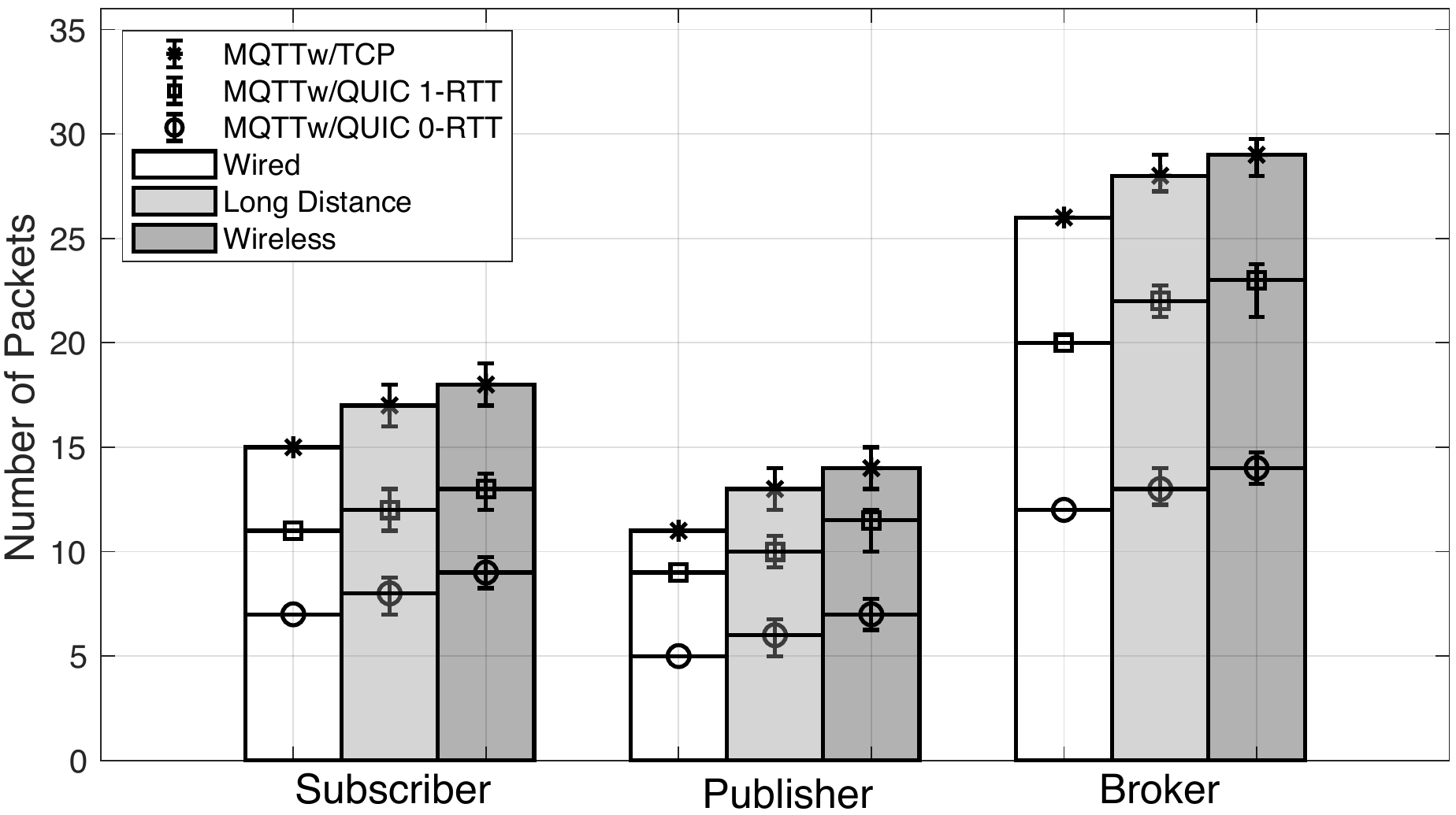}
    \captionsetup{font=scriptsize}
    \caption{The number of packets exchanged during the connection establishment phase in the wired, wireless, and long-distance testbeds. The error bars show the lower quartile and higher quartile of the collected results.}
    \label{fig:NumofPackets}
\end{figure}

This section evaluates the number of packets exchanged between devices when using MQTTw/TCP and MQTTw/QUIC during the connection establishment process.
Before presenting the results, we first study the sequence of packet exchanges between a client and a server.
Figures \ref{fig:conn_pkt_exch}(a), (b), and (c) show this sequence in MQTTw/TCP and MQTTw/QUIC's 1-RTT, and 0-RTT, respectively.
Please note that these figures demonstrate ideal scenarios when there is no packet drop. 
In addition, MQTT ping packets are excluded to simplify the evaluations.
Figure \ref{fig:NumofPackets} and Table \ref{num_of_packets_result} summarize the results from the subscriber, publisher and broker point of views using the three testbed types mentioned earlier.

\begin{table*}[ht]
\centering
\caption{Performance improvement of MQTTw/QUIC versus MQTTw/TCP in terms of the number of packets exchanged during the connection establishment phase.} 
\label{num_of_packets_result}
\begin{tabular}{|c|c||c|c|c|c|c|c|c|c|c|}
\hline
\rowcolor[HTML]{FFFFFF} 
\multicolumn{2}{|c||}{\cellcolor[HTML]{FFFFFF}{\color[HTML]{000000} Testbed}} & \multicolumn{3}{c|}{\cellcolor[HTML]{FFFFFF}{\color[HTML]{000000} \textbf{Wired}}} & \multicolumn{3}{c|}{\cellcolor[HTML]{FFFFFF}{\color[HTML]{000000} \textbf{Wireless}}} & \multicolumn{3}{c|}{\cellcolor[HTML]{FFFFFF}{\color[HTML]{000000} \textbf{Long Distance}}} \\ \hline \hline
\rowcolor[HTML]{FFFFFF} 
\multicolumn{2}{|c||}{\cellcolor[HTML]{FFFFFF}{\color[HTML]{000000} Device}} & \multicolumn{1}{l|}{\cellcolor[HTML]{FFFFFF}{\color[HTML]{000000} Subscriber}} & \multicolumn{1}{l|}{\cellcolor[HTML]{FFFFFF}{\color[HTML]{000000} Publisher}} & \multicolumn{1}{l|}{\cellcolor[HTML]{FFFFFF}{\color[HTML]{000000} Broker}} & \multicolumn{1}{l|}{\cellcolor[HTML]{FFFFFF}{\color[HTML]{000000} Subscriber}} & \multicolumn{1}{l|}{\cellcolor[HTML]{FFFFFF}{\color[HTML]{000000} Publisher}} & \multicolumn{1}{l|}{\cellcolor[HTML]{FFFFFF}{\color[HTML]{000000} Broker}} & \multicolumn{1}{l|}{\cellcolor[HTML]{FFFFFF}{\color[HTML]{000000} Subscriber}} & \multicolumn{1}{l|}{\cellcolor[HTML]{FFFFFF}{\color[HTML]{000000} Publisher}} & {\color[HTML]{000000} Broker} \\ \hline
\cellcolor[HTML]{FFFFFF}{\color[HTML]{000000} } & \cellcolor[HTML]{FFFFFF}{\color[HTML]{000000} 1-RTT} & {\color[HTML]{000000} 36.84\%} & {\color[HTML]{000000} 33.33\%} & {\color[HTML]{000000} 35.29\%} & {\color[HTML]{000000} 48.18\%} & {\color[HTML]{000000} 42.85\%} & {\color[HTML]{000000} 47.91\%} & {\color[HTML]{000000} 45.83\%} & {\color[HTML]{000000} 35.29\%} & {\color[HTML]{000000} 40\%} \\ \cline{2-11} 
\multirow{-2}{*}{\cellcolor[HTML]{FFFFFF}{\color[HTML]{000000} \begin{tabular}[c]{@{}c@{}}Improvement \\ vs MQTTw/TCP\end{tabular}}} & \cellcolor[HTML]{FFFFFF}{\color[HTML]{000000} 0-RTT} & {\color[HTML]{000000} 47.36\%} & {\color[HTML]{000000} 46.66\%} & {\color[HTML]{000000} 47.05\%} & {\color[HTML]{000000} 55.55\%} & {\color[HTML]{000000} 52.38\%} & {\color[HTML]{000000} 56.25\%} & {\color[HTML]{000000} 54.16\%} & {\color[HTML]{000000} 47.05\%} & {\color[HTML]{000000} 50\%} \\ \hline
\end{tabular}
\end{table*}

As the results show, MQTTw/QUIC reduces the number of packets exchanged with the broker by up to 56.25\%.
In addition, the wireless testbed shows the highest performance improvement compared to the other two testbeds, which is due to its higher packet loss rate.
Therefore, since the number of packets required by MQTTw/QUIC is lower in this scenario, compared to MQTTw/TCP, the total number of packet re-transmissions is lower.
Compared to the wired testbed, the long-distance testbed shows a higher performance improvement.
In the wired network, all the three devices are in same broadcast domain, therefore, no layer-3 routing is performed.
This an ideal situation since all the devices are connected directly through a LAN and the probability of packet loss is almost zero.
However, packets might be lost or significantly delayed in the long-distance testbed due to events such as congestion and multipath.
As IoT applications such as smart homes usually include lossy links, these results indicate the significant potential benefits of using QUIC in these scenarios.


\subsection{Head-of-Line Blocking}
\label{head-of-line}
The head-of-line blocking problem occurs when the receiver is waiting for the dropped packets to be re-transmitted in order to complete the packet reordering before delivering them to the application.
To simulate this problem in our testbeds, a FreeBSD \cite{mckusick2014design} router was introduced in between the publisher and broker. 
FreeBSD enables us to intercept packets based on the firewall rules and their network buffers. 
During this experiment, which ran for 300 seconds, data packets are intercepted based on their flow ID. 
The flow ID is a unique tuple of source IP address and source port to identify the connection. 
The FreeBSD's firewall function \texttt{ipfw\_chk} is used to intercept the packets and drop them.
In order to replicate a real-world scenario, data packets were dropped randomly at fixed intervals. 
For instance, to simulate a 10{\%} drop scenario, every 10th packet belonging to the same ID is dropped.
In these experiments, latency is computed as the interval between the instance a packet leaves the publisher until the reception of that packet by the subscriber.
To accurately measure latency without introducing extra traffic, WiringPi \cite{henderson2013wiring} has been integrated into our testbed. 
For each received message, the receiver uses the WiringPi library to notify the sender by generating a signal to toggle a pin connected to the sender. 
\begin{table*}[t!]
\caption{Performance improvement of MQTTw/QUIC versus MQTTw/TCP in terms of packet delivery latency in the presence of packet drops. The drop rate is changed between 10\% to 50\%.}
\centering
\setlength{\tabcolsep}{5pt}
\label{HOL_latency}
\begin{tabular}{|
>{\columncolor[HTML]{FFFFFF}}c||c|c|c|c|c|c|c|c|c|}
\hline

Testbed & \multicolumn{3}{c|}{\cellcolor[HTML]{FFFFFF}\textbf{Wired}} & \multicolumn{3}{c|}{\cellcolor[HTML]{FFFFFF}\textbf{Wireless}} & \multicolumn{3}{c|}{\cellcolor[HTML]{FFFFFF}{\color[HTML]{000000} \textbf{Long Distance}}} \\ \hline
\hline
Drop Rate & 10\% & 20\% & 50\% & 10\% & 20\% & 50\% & 10\% & 20\% & 50\% \\ \hline
\begin{tabular}[c]{@{}c@{}}Latency of \\MQTTw/TCP\end{tabular} & 3.114 ms & 5.234 ms & 10.6904 ms & 11.2176 ms & 15.7451 ms & 21.4137 ms & 44.3578 ms & 54.9340 ms & 75.6283 ms \\ \hline
\begin{tabular}[c]{@{}c@{}}Latency of \\MQTTw/QUIC\end{tabular}  & 1.3833 ms & 3.5680 ms & 8.9389 ms & 7.1099 ms & 10.5313 ms & 18.5313 ms & 33.0088 ms & 44.0462 ms & 64.3635 ms \\ \hline
\begin{tabular}[c]{@{}c@{}}Improvement\\vs MQTTw/TCP\\ \end{tabular} & {\color[HTML]{000000} 55.57\%} & {\color[HTML]{000000} 31.83\%} & \multicolumn{1}{c|}{{\color[HTML]{000000} 16.38\%}} & \multicolumn{1}{c|}{{\color[HTML]{000000} 36.61 \%}} & \multicolumn{1}{c|}{{\color[HTML]{000000} 33.11\%}} & \multicolumn{1}{c|}{{\color[HTML]{000000} 13.46\%}} & \multicolumn{1}{c|}{{\color[HTML]{000000} 25.58\%}} & \multicolumn{1}{c|}{{\color[HTML]{000000} 19.81\%}} & \multicolumn{1}{c|}{{\color[HTML]{000000} 14.89\%}} \\ \hline
\end{tabular}
\end{table*}

Table \ref{HOL_latency} summarizes the results for all the three testbeds. 
Dropping random packets intermittently forces the sender to retransmit the lost packets. 
As the results show, TCP's latency is higher than QUIC's in all the scenarios.
This is due to the packet re-ordering delay in TCP, which happens when a line of packets is being held up by the receiver when prior packets are lost.
In contrast, QUIC is based on UDP, which does not hold up the received packets while the lost packets are being re-transmitted. 
This enables the receiver's application layer to perform the decryption operation as soon as packets arrive.

In the wireless and long distance testbeds, apart from the intentional drops introduced, both MQTTw/QUIC and MQTTw/TCP show higher latencies as they suffer from the additional packet drops, compared to the wired testbed.
Therefore, the improvement margins between MQTTw/TCP and MQTTw/QUIC in wireless and long distance testbeds are lower than the wired network. 
For example, for the 10\% drop rate scenario, MQTTw/QUIC reported 1.3833 ms and 7.1099 ms latencies in the wired and wireless testbeds, respectively. 
Therefore, the highest improvement achieved is for wired networks.

\subsection{TCP Half-Open Connections}
\label{tcpHalfOpen}
An act of receiving data in TCP is passive, which means that a dropped connection can only be detected by the sender and not by the receiver. 
In order to simulate the TCP half-open connection problem, 100 connections were established by 10 publishers using different topics and client IDs. 
One message is transmitted per second over each connection.
To generate TCP half-open connections, the keep-alive message transmission mechanism of MQTT has been disabled. 
The half-open connections are detected using Linux command \texttt{ss -a}. 
In order to reveal the overhead of half-open connections, memory usage and processor utilization level were measured using the Linux's \texttt{top} utility.
To generate half-open connections, all the publishers are restarted in the middle of a message flow. 
In this scenario, the broker is unaware of the connection tear-down.

\begin{figure}[t]
   \centering
    \includegraphics[width=0.82 \linewidth]{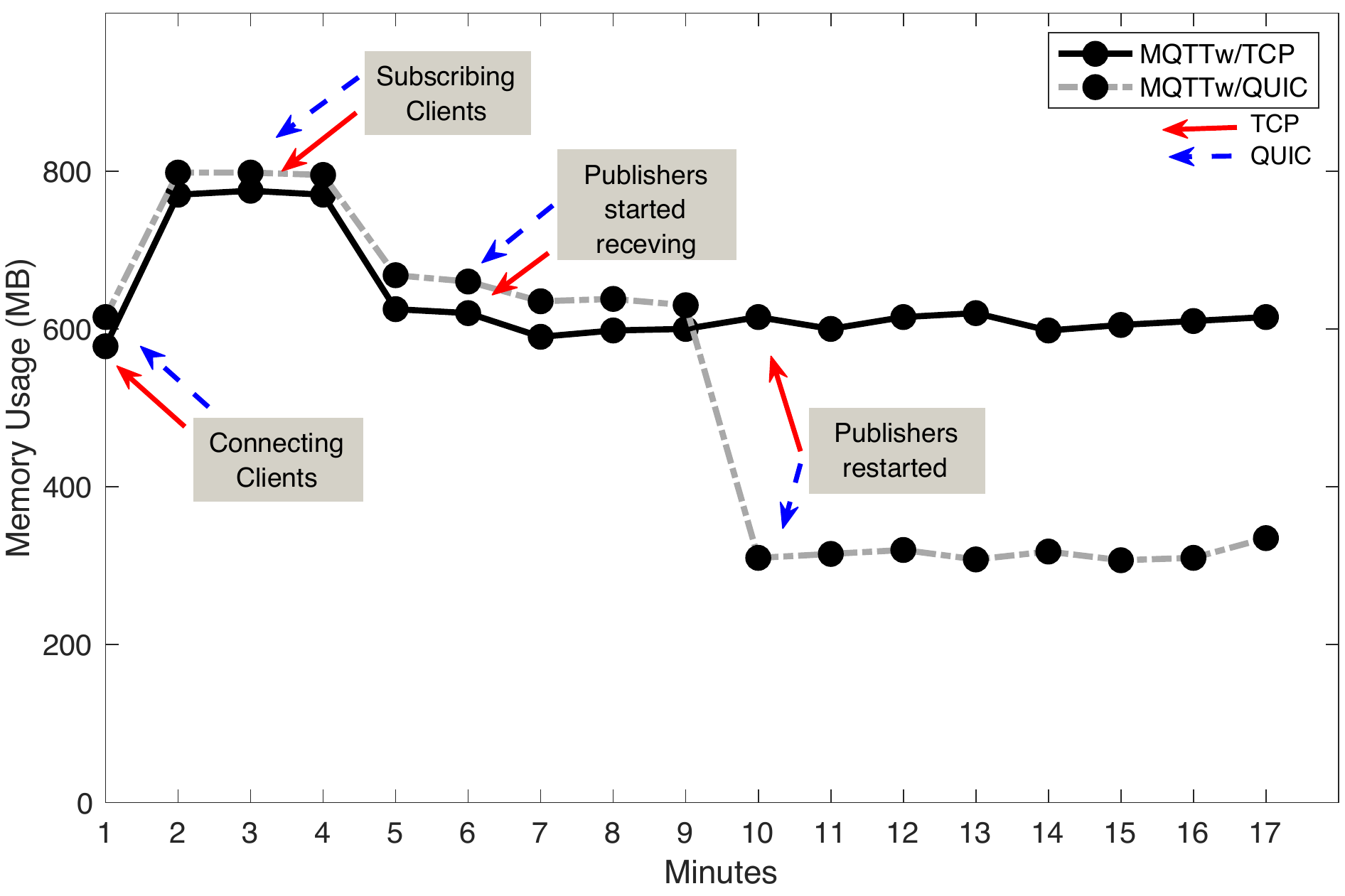}
    \captionsetup{font=scriptsize}
    \caption{The memory utilization of broker. 
    While the memory utilization level of MQTTw/QUIC is dropped after the publishers were restarted, the memory utilization of MQTTw/TCP remains high.}
    \label{fig:tcp_half_open_memory_usage}
\end{figure}

\begin{figure}[t]
   \centering
    \includegraphics[width=0.83 \linewidth]{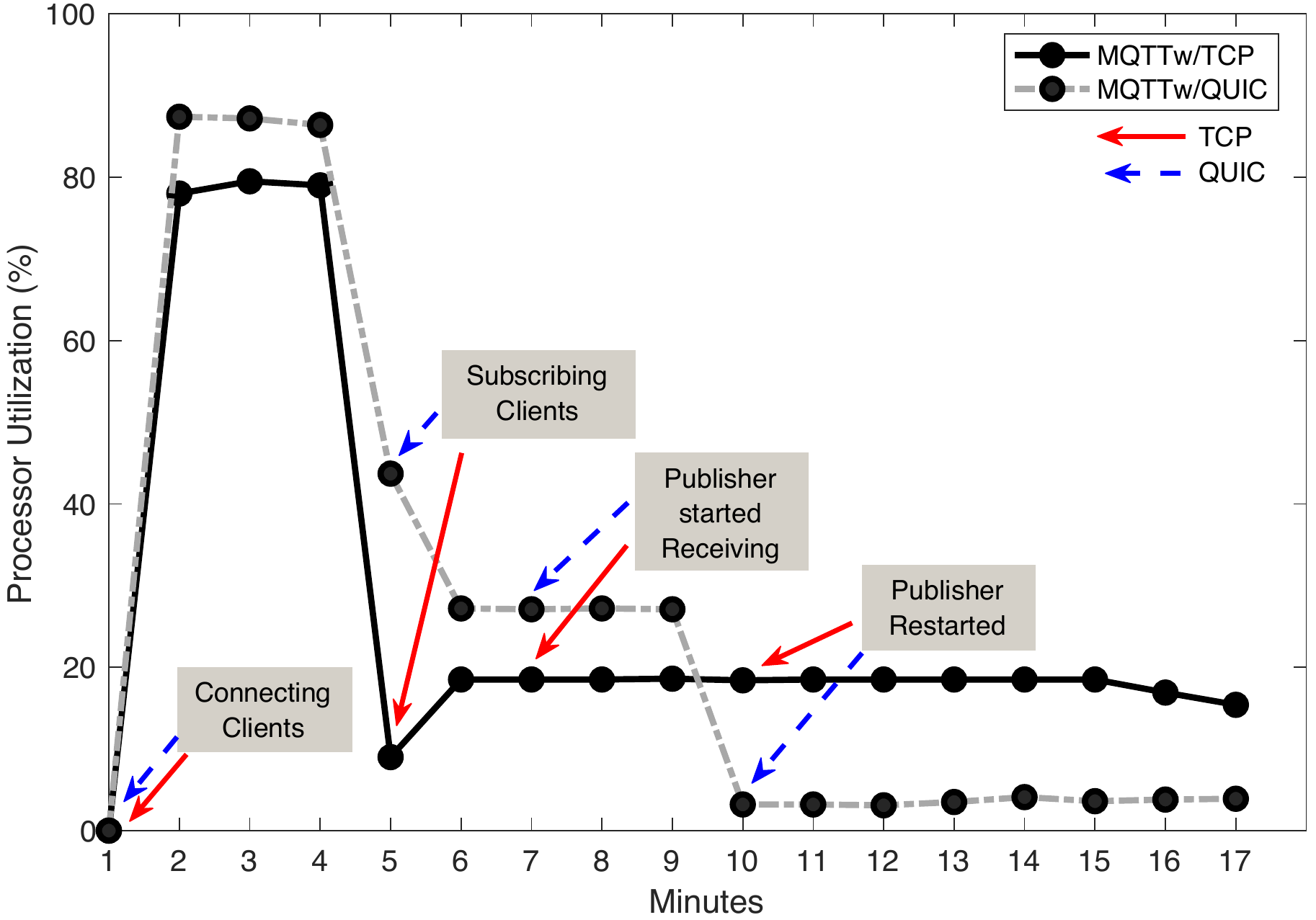}
    \captionsetup{font=scriptsize}
    \caption{The processor utilization of broker. 
    Although the processor utilization of MQTTw/QUIC is more than MQTTw/TCP, after the publishers were restarted, MQTTw/QUIC shows a higher drop in processor usage as QUIC runs over UDP and does not require any connection state information.}
    \label{fig:tcp_half_open_CPU_usage}
\end{figure}

Figures \ref{fig:tcp_half_open_memory_usage} and  \ref{fig:tcp_half_open_CPU_usage} show the resource utilization of the broker.
Both figures present the results of a single experiment to reveal the variations of memory and processor utilization versus time.
The duration of this experiment is long enough to include the process of client connection, subscription, publishing, and restart of the clients.
After each of the aforementioned operations, we wait for the processor and memory utilization to become stable and then start the next operation.
These results shows that MQTTw/QUIC releases the resources as soon as it detects the connection is broken. 
Specifically, the MQTTw/QUIC's UDP socket times out under 1 minute and QUIC starts its draining period.
On the other hand, TCP connections still wait for incoming packets, as there is no mechanism to detect whether the connections opened by the publishers are still active or not. 
In summary, it is observed that MQTTw/QUIC reduces processor utilization between 74.67\% to 83.24\% and lowers memory utilization between 45.52\% to 50.32\%, compared to MQTTw/TCP.
It must be noted that MQTTw/QUIC achieves these improvements without introducing any communication overheads such as keep-alive packets.

\subsection{Connection Migration} 
\label{connection_migration_result}
As mentioned earlier, connection reestablishment due to roaming imposes energy overhead.
This section evaluates the efficiency of the proposed system during connection migrations.
To this end, 100 connections were established from 10 publishers to a broker, and the subscribers were subscribed to all the topics. 
The \texttt{nload} utility was used to measure throughput.
In order to emulate the change of network, the IP address of the broker's interface connected to the subscriber is changed every 5 minutes.
The total duration of the experiment is 16 minutes to include three connection reestablishment triggers and allow the publishers and broker to stabilize after each migration.

\begin{figure}[t]
   \centering
    \includegraphics[width=0.82 \linewidth]{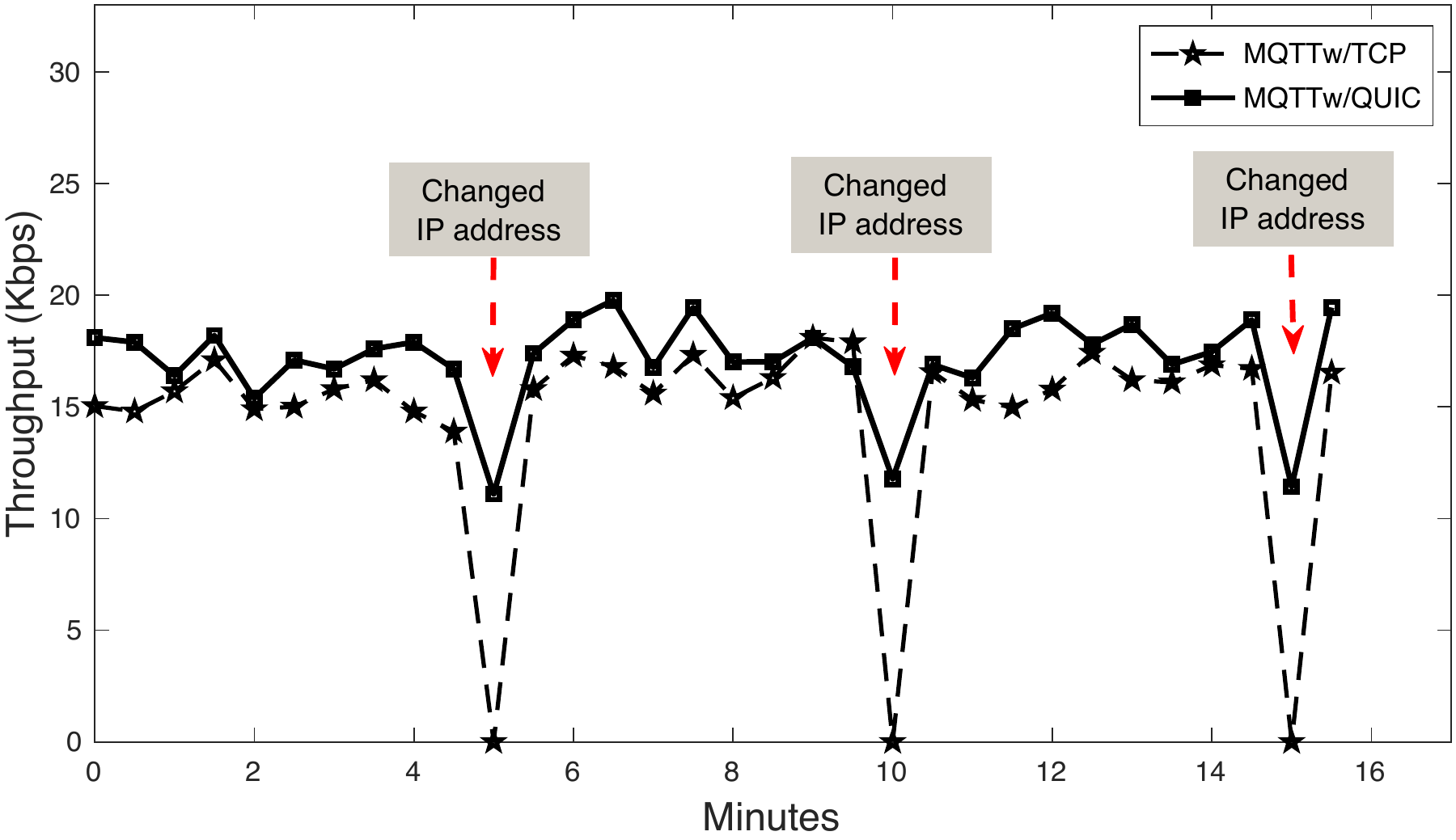}
    \captionsetup{font=scriptsize}
    \caption{The effect of connection migration on throughput.
     After each IP renewal, TCP re-establishes the connections.
     In contrast, QUIC connections are survived and only a slight throughput degradation can be observed due to the change in interface IP address.}
    \label{fig:Connection_Migration}
\end{figure}

Figure \ref{fig:Connection_Migration} shows the results.
Whenever a change in the connection parameters occurs, MQTTw/TCP has to repeat the cumbersome process of connection re-establishment.
In this case, its bandwidth drops to zero and picks up again after the connection re-establishment process completes. 
In contrast, MQTTw/QUIC's throughput shows a slight reduction due to the change in the interface state, but connections were migrated to the new IP unscathed.
Specifically, QUIC allows the end point devices to survive in such events, requiring only to have a stable IP address during the handshake process to obtain the 1-RTT keys.
This feature is particularly useful in mobile IoT scenarios such as medical and industrial applications \cite{dezfouli2017rewimo}.
Another useful case is to make the clients resilient against NAT rebinding.

\section{Related Work} 
\label{relatedwork}
This section is composed of two sub-sections. 
The first sub-section reviews the IoT application layer protocols and their employed transport layer protocols.
The second sub-section reviews the literature relevant to QUIC and compares the features offered by the proposed MQTTw/QUIC implementation against existing IoT application layer protocols.

\subsection{IoT Application Layer Protocols}
This section overviews the main IoT application layer protocols and their adoption of transport layer protocols.

\subsubsection{MQTT (Message Queuing Telemetry Transport)}
MQTT is a lightweight IoT application protocol for machine-to-machine connectivity.
The protocol specification \cite{banks2014message} mandates the use of a connection-oriented transport protocol to ensure packet ordering and reliable delivery. 
Therefore, most implementations \cite{Paho_Eclipse, hiveMQ, CocoaMQTT, MosquittoIOsphere, HaskellMQTT} use TCP/IP with TLS/SSL to satify this requirement. 
In some cases \cite{SCTPMQTT}, the Stream Control Transmission Protocol (SCTP), an alternative to TCP, has been used as well.
However, the key problems with SCTP when used in IoT domains are: 
(i) 4-way handshake, which increases latency compared to the TCP's 3-way handshake. 
(ii) SCTP is mainly used for multihoming and redundancy. Whereas in IoT, redundancy is not an essential aspect compared to latency. 
(iii) SCTP is not suitable for heterogeneous networks \cite{fu2004sctp}. 
The authors in \cite{stewart2004sctp} studied TCP, UDP and SCTP, and justified why SCTP is not the best fit as an IoT protocol.

\subsubsection{XMPP (Extensible Messaging and Presence Protocol)} 
XMPP offers both publish/subscribe (asynchronous) and request/response (synchronous) messaging capabilities. 
This protocol uses TLS/SSL for security, and relies on TCP to ensure reliability.
XMPP is most suited for real-time applications, with small message footprints and low latency in message exchange \cite{bendel2013service}.
Although XMPP is preferred over CoAP in data-centric IoT scenarios due to offering the publish/subscribe mechanism, the lack of QoS provisioning might not be acceptable in mission-ciritical applications.
Furthermore, its underlying transport protocol, TCP, inherits the shortcomings we discussed earlier.
Google has reported various XMPP incompatibilities with their new applications\cite{GoogleXMPP} and declared it as an obsolete protocol.

\subsubsection{REST (Representational State Transfer)} 
REST is an architectural paradigm.
REST uses the HTTP \cite{fielding2014rfc} methods (GET, POST, PUT and DELETE) to provide a resource-oriented messaging system. 
All actions are performed by simple request/response (synchronous) HTTP commands. 
It has a built-in accept HTTP header to determine the format of the data, which is usually JSON (JavaScript Object Notation) or XML (Extensible Markup Language). 
REST is easy to implement and is supported by most M2M cloud platforms. 
Other features supported by REST are caching, authentication and content type negotiation \cite{upadhyaya2011migration}.
This protocol, however, does not offer high energy efficiency since polling must be performed in order to check the availability of data.

REST provides a one-way connection between the client and server. 
Client only connects to the server when it has to either push or pull the data. 
On the other hand, MQTT relies on a two-way established connection between the server and client.
This enables the server to respond to client's request instantly \cite{RESTvsMQTT}.
In contrast, since REST is a request/reply protocol, it does not need to establish a long-term connection.
However, in lossy networks, in particular, establishing connections repeatedly increases energy consumption \cite{savolainen2014measuring}. 
BevyWise Networks conducted experiments and estimated that their proprietary \textit{MQTTRoute} broker \cite{MQTTRoute} consumes 20\% less power than REST. 
Their findings concluded that the primary reason behind more power consumption in REST was due to the resources used for connecting, re-connecting, and cleaning up both the server and client connection states. 
As another example, \cite{RESTvsMQTT} shows that MQTT is up to 25 times faster than REST in terms of data transfer rate.

The availability of a REST server might be limited if the devices are behind a firewall. 
A common scenario of firewall deployment is when all new incoming connections are blocked, while outgoing connections are always allowed due to zoning. 
In this case, establishing connection to IoT devices would not be possible.
MQTT solves this issue by relying on two-way connections. 

\subsubsection{AMQP (Advanced Message Queuing Protocol)} 
Although AMQP is agnostic to transport protocol, it uses TCP (and TLS) by default.
AMPQ offers a publish/subscribe model for messaging.
This protocol provides a reliable connection by utilizing a \textit{store and forward} mechanism \cite{bhimani2018message,johnsen2013evaluation}, thereby offering reliability in the presence of network disruptions. 
The authors in \cite{fernandes2013performance} show that AMQP can send a larger amount of messages per second, compared to REST. 
Studies show that AMQP has a low success rate in low bandwidth, but the success rate increases as bandwidth increases \cite{luzuriaga2015comparative,johnsen2013evaluation}.



\subsubsection{MQTT-SN (Message Queuing Telemetry Transport - Sensor Networks)} 
MQTT-SN is a variant of MQTT, which is mainly designed for very resource-constrained devices.
Specifically, this protocol aims to reduce the high energy consumption and bandwidth usage of MQTT networks \cite{stanford2013mqtt}. 
Although MQTT-SN can use UDP as its underlying transport protocol, it is agnostic to the underlying transport services.
Apart from introducing UDP, MQTT-SN also mandates the inclusion of a gateway (GW) and forwarder. 
The primary function of the gateway is to transform UDP packets into standard MQTT packets and transmit the packets to a broker.
There are two categories for this connection: transparent and aggregating. 
Using a transparent gateway, every MQTT-SN node owns a dedicated TCP connection to the broker. 
On the other hand, an aggregating gateway only has one TCP connection to the broker, which is then used in tunneling. 
The forwarder receives MQTT-SN frames on the wireless side, and encapsulates and forwards them to the gateway.
For the return traffic, frames are decapsulated and sent to the clients without applying any modifications.
MQTT-SN avails DTLS and TLS to provide security for UDP and TCP, respectively.
Using a gateway is mandatory in MQTT-SN. 
The authors in \cite{amaran2015comparison} proved MQTT-SN performs 30\% faster than CoAP. 

\subsubsection{CoAP (Constrained Application Protocol)}
CoAP eliminates the overhead of TCP by relying on UDP \cite{keoh2013profiling,keoh2014securing}, and DTLS is used to secure the connections.
CoAP implements reliability by defining two bits in each packet header.
These bits determine the type of message and the required Quality of Service (QoS) level. 
The authors in \cite{thangavel2014performance} show that the delay of CoAP is higher than MQTT, which is caused by packet losses when UDP is replaced with TCP.


\subsection{QUIC} 
Various studies have evaluated the performance of QUIC in the context of Internet traffic.
Google has deployed this protocol in its front-end servers that collectively handle billions of requests per day \cite{langley2017quic}.
Google claims that QUIC outperforms TCP in a variety of scenarios such as reducing latency by 8\% in Google search for desktop users and 3.6\% for mobile users. 
In video streaming, latency in YouTube playbacks is reduced by 18\% for desktop and 15.3\% for mobile users. 
Currently, QUIC is used for 30\% of Google's egress traffic. 

\begin{table*}[htp]
\begin{center}
\begin{threeparttable}
\caption{Comparison of IoT application layer protocols}
\label{result}
\begin{tabular}{|
>{\columncolor[HTML]{FFFFFF}}c ||
>{\columncolor[HTML]{FFFFFF}}c |
>{\columncolor[HTML]{FFFFFF}}c |
>{\columncolor[HTML]{FFFFFF}}c |
>{\columncolor[HTML]{FFFFFF}}c |
>{\columncolor[HTML]{FFFFFF}}c |
>{\columncolor[HTML]{FFFFFF}}c |
>{\columncolor[HTML]{C0C0C0}}c |}
\hline
                                                                                Protocol & CoAP & MQTT & MQTT-SN & XMPP & REST & AMQP & MQTTw/QUIC \\ \hline \hline
UDP Compatible  & \multicolumn{1}{c|}{$\surd{}$}     & \multicolumn{1}{c|}{X}     &  \multicolumn{1}{c|}{$\surd{}$}       & \multicolumn{1}{c|}{X}     & \multicolumn{1}{c|}{X}     & \multicolumn{1}{c|}{$\surd{}$}     & \multicolumn{1}{c|}{\cellcolor[HTML]{C0C0C0}$\surd{}$}             \\ \hline
TCP Compatible  & \multicolumn{1}{c|}{X}     & \multicolumn{1}{c|}{$\surd{}$}     &  \multicolumn{1}{c|}{$\surd{}$}       & \multicolumn{1}{c|}{$\surd{}$}     & \multicolumn{1}{c|}{$\surd{}$}     & \multicolumn{1}{c|}{$\surd{}$}     & \multicolumn{1}{c|}{\cellcolor[HTML]{C0C0C0}X}             \\ \hline
Multiplexing Capability                                                                       & \multicolumn{1}{c|}{X}     & \multicolumn{1}{c|}{X}     & \multicolumn{1}{c|}{X}        & \multicolumn{1}{c|}{X}     & \multicolumn{1}{c|}{X}      & \multicolumn{1}{c|}{X}      & \multicolumn{1}{c|}{\cellcolor[HTML]{C0C0C0}$\surd{}$}              \\ \hline
0-RTT Capable   & \multicolumn{1}{c|}{X}      & \multicolumn{1}{c|}{X}     & \multicolumn{1}{c|}{X}        & \multicolumn{1}{c|}{X}     & \multicolumn{1}{c|}{X}     & \multicolumn{1}{c|}{X}     & \multicolumn{1}{c|}{\cellcolor[HTML]{C0C0C0}$\surd{}$}             \\ \hline
Fixing Head-of-Line Blocking      & N/A & \multicolumn{1}{c|}{X} & \multicolumn{1}{c|}{X} & \multicolumn{1}{c|}{X} & \multicolumn{1}{c|}{X} & \multicolumn{1}{c|}{X} & \multicolumn{1}{c|}{\cellcolor[HTML]{C0C0C0}$\surd{}$} \\ \hline

\begin{tabular}[c]{@{}l@{}} Fixing TCP Half-Open Problem \\(Adaptability to Lossy Networks) \end{tabular} & \multicolumn{1}{c|}{N/A}     & \multicolumn{1}{c|}{X}     & \multicolumn{1}{c|}{X}        & \multicolumn{1}{c|}{X}     & \multicolumn{1}{c|}{X}     & \multicolumn{1}{c|}{X}     & \multicolumn{1}{c|}{\cellcolor[HTML]{C0C0C0}$\surd{}$}             \\ \hline
Supporting Connection Migration                    & \multicolumn{1}{c|}{X}     & \multicolumn{1}{c|}{X}     &  \multicolumn{1}{c|}{X}       & \multicolumn{1}{c|}{X}     & \multicolumn{1}{c|}{X}     & \multicolumn{1}{c|}{X}     & \multicolumn{1}{c|}{\cellcolor[HTML]{C0C0C0}$\surd{}$}             \\ \hline
\end{tabular}
\label{comparison_all}
\end{threeparttable}
\end{center}
\end{table*}

In \cite{qian2018achieving}, the authors evaluate the performance of multiple QUIC streams in LTE and WiFi networks.
Their experiments show that for mobile web page, median and 95th percentile completion time can be improved by up to 59.1\% and 72.3\%, respectively, compared to HTTP.
The authors in \cite{kakhki2017taking} show that QUIC outperforms TCP with respect to Page Load Time (PLT) in different desktop user network conditions such as added delay and loss, variable bandwidth, and mobile environment. 
The studies of \cite{cook2017quic} show that QUIC outperforms TLS when used in unstable networks such as WiFi \cite{sathiaseelan2013internet}. 
Specifically, the authors performed experiments to measure PLT for YouTube traffic with different delays (0ms, 50ms, 100ms and 200ms). 
These values were adopted from \cite{AkamaiQUIC} for QUIC to simulate real networks. 
Assuming the PLT is $x$ when the introduced delay is 0, these observations have been made.
For the second connection, the PLT is doubled, and it is increased by 400ms for each consecutive connection.
However, for HTTP-2 repeat connections, the PLT is multiplied by the connection number for each consecutive connection, i.e., $2x$, $3x$, etc.
In another study \cite{megyesi2016quick}, the authors showed that in more than 40\% of scenarios the PLT of QUIC is lower than SPDY \cite{belshe2012spdy} and TLS combined.

The authors in \cite{bakri2015http} have evaluated the performance of HTTP2 with QUIC for Multi-User Virtual World (MUVW) and 3D web.
MUVW networks mostly run on UDP \cite{oliver2009traffic,oliver2010virtual}, and require the network administrator to open 50 or more UDP ports on the firewall. 
Since these ports carry UDP traffic, an application protocol capable of offering connection-orientated streams and security is required.
QUIC is an ideal candidate to fill this void.

The implementation and integration of QUIC with MQTT are complementary to the aforementioned studies since none of them have considered the applicability of QUIC in IoT applications.
Specifically, this work reveals the importance and benefits of integrating MQTT with QUIC when used in various types of IoT networks including local, wireless and long-distance networks.
The results presented in this paper, in particular, confirm 56\% lower packet exchange overhead, 83\% lower processor utilization, 50\% lower memory utilization, and 55\% shorter delivery delay of MQTTw/QUIC compared to MQTTw/TCP.
Table \ref{comparison_all} compares MQTTw/QUIC versus the existing application layer protocols.

\section{Conclusion}
\label{conclusion}
Enhancing transport layer protocols is essential to ensure compatibility while addressing the particular demands of IoT networks.
Specifically, shifting the implementation of protocols to the user space brings substantial benefits such as offloading the cost of modifying the kernel and enhancing processing speed.
In addition, given the resource-constrained nature of IoT devices, reducing communication overhead is essential. 
Unfortunately, these requirements are not satisfied by TCP/TLS and UDP/DTLS.

In this paper, we justified the potential benefits of QUIC compared to TCP/TLS and UDP/DTLS in IoT scenarios and presented its integration with MQTT protocol.
Specifically, we showed the software architecture proposed as well as the agents developed to enable the communication between MQTT and QUIC.
Three different testbeds were used to evaluate the performance of MQTTw/QUIC versus MQTTw/TCP.
The results confirmed that MQTTw/QUIC reduces connection establishment overhead, lowers delivery latency, reduces processor and memory utilization, and shortens the level and duration of throughput degradation during connection migration significantly compared to MQTTw/TCP.

Some of the potential areas of future work are as follows:
First, the processor utilization of QUIC is higher than TCP. 
This is due to the cost of encryption and packet processing while maintaining QUIC's internal state. 
To reduce the overhead of encryption, for example, the ChaCha20 optimization technique could be used \cite{langley2016chacha20}.
Although packet processing cost was minimized by using asynchronous packet reception in the kernel by applying a memory-mapped ring buffer, a.k.a., \texttt{PACKET\_RX\_RING}, the cost is still higher than TCP. 
To minimize this problem, one solution is to employ \textit{kernel bypass} \cite{rizzo2012netmap, paolino2015snabbswitch, intel2014data} to bring packet processing into the user space. 
Second, the core functionality of QUIC assumes a maximum MTU of 1392 bytes for handshake packets \cite{QUICMtuProb}. 
This includes 14 bytes for Ethernet header, 20 bytes for IP header, 8 bytes for UDP, and 1350 bytes for QUIC. 
At present, the 1392 bytes is a static value in the client side, which is based on observational testing.
All handshake packets are required to be padded to the full size in both directions. 
This limitation prohibits IP fragmentation and as a result limits the path MTU discovery.
Third, as reviewed in this paper, QUIC offers many features that can be employed to further enhance the peroformance of IoT networks.
For example, since most IoT applications do not rely on high throughput, we did not evaluate the effect of QUIC's flow control.
However, the study of this mechanism is left as a future work, which can be justified by IoT applications such as motion detection and image classification.

\ifCLASSOPTIONcaptionsoff
  \newpage
\fi

\bibliographystyle{IEEEtran}
\bibliography{references}

\end{document}